\documentclass[superscriptaddress,aps,10pt]{revtex4}

\usepackage[dvips]{graphics}
\usepackage{graphicx}
\usepackage{amsfonts}
\usepackage{amssymb}
\usepackage{amsmath}
\usepackage{color}

\begin{document}

\title{Symmetry-breaking Effects for
Polariton Condensates in Double-Well Potentials}

\author{A.S.\ Rodrigues}
\affiliation{Departamento de F\'{\i}sica/CFP, Faculdade de Ci\^{e}ncias,
Universidade do Porto, R. Campo Alegre, 687 - 4169-007 Porto, Portugal}

\author{P.G.\ Kevrekidis}
\affiliation{Department of Mathematics and Statistics, University of Massachusetts,
Amherst MA 01003-4515, USA}

\author{J.\ Cuevas}
\affiliation{Grupo de F\'{\i}sica No Lineal.  Departamento de F\'{\i}sica Aplicada I.
Escuela Polit\'ecnica Superior, Universidad de Sevilla, C/ Virgen de \'Africa, 7, 41011-Sevilla, Spain}

\author{R.\ Carretero-Gonz\'alez}
\affiliation{Nonlinear Dynamical Systems Group,%
\footnote{\texttt{URL}: http://nlds.sdsu.edu}%
Department of Mathematics and Statistics, and Computational Science
Research Center, San Diego State University, San Diego CA, 92182-7720, USA}

\author{D.J.\ Frantzeskakis}
\affiliation{Department of Physics, University of Athens, Panepistimiopolis,
Zografos, Athens 157 84, Greece}

\begin{abstract}
We study the existence, stability, and dynamics of symmetric and
anti-symmetric states of quasi-one-dimensional polariton condensates in
double-well potentials, in the presence of nonresonant pumping and nonlinear damping.
Some prototypical features of the system, such as the
bifurcation of asymmetric solutions, are similar to
the Hamiltonian analog of the double-well system
considered in the realm of atomic condensates. Nevertheless,
there are also some nontrivial differences including, e.g., the
unstable nature of both the parent and the daughter branch emerging
in the relevant pitchfork bifurcation for slightly larger
values of atom numbers. Another interesting feature that does not
appear in the atomic condensate case is that the bifurcation for attractive
interactions is slightly sub-critical instead of supercritical.
These
conclusions of the bifurcation analysis are corroborated by direct
numerical simulations examining the dynamics of the system in
the unstable regime.
\end{abstract}

\date{\today}

\maketitle

\section{Introduction}

Over the past few years, a novel direction in the study of
Bose-Einstein condensation has captured a considerable amount of
attention. This concerns the observation of exciton-polariton Bose-Einstein condensates (BECs)
in semiconductor microcavities \cite{kasp1_and_more,balili,lai,deng}. A fundamental
feature of these exciton-polariton BECs is that, upon confinement, the
excitons (bound pairs of electrons and holes) couple strongly to the
incident light creating the polariton
quasi-particles~\cite{micro_cavity_polaritons}. The resulting exciton-polariton
BEC possesses a number of remarkable properties that we briefly touch upon
below.

The radiative lifetime
of the polaritons is the shorter relaxation time scale of the system being
of the order of 1--10~ps~\cite{benoit}. On the other hand, the light mass
of the exciton-polaritons provides this system with a significantly higher
condensation temperature.
The photonic component of
the exciton-polaritons is responsible for their short lifetime which, in turn,
does not allow thermalization;
instead, it produces a non-equilibrium
condensate, wherein the presence of external pumping from an exciton
reservoir is critical towards a counter-balance of the polariton loss.
In such genuinely non-equilibrium condensates, numerous remarkable
features have been not only theoretically predicted but also
experimentally established;
these include the flow without scattering (analog of the
flow without friction) \cite{amo1}, the existence of vortices \cite{lagou1}
(see also Ref.~\cite{roumpos} for vortex dipole dynamics and Ref.~\cite{roumpos1}
for observations thereof),
the collective dynamics \cite{amo2}, as well as remarkable applications
such as spin switches \cite{amo3} and light emitting diodes \cite{amo4}
operating even near room temperatures.

Perhaps the most customary approach to modeling exciton-polariton BECs
involves the coupling of the evolution of the polaritons to that
of the exciton reservoir which enables their production (and which features
diffusive spatial dynamics of the excitons); this way, the model takes the form of two
coupled complex Ginzburg-Landau (cGL) equations describing the evolution of exciton 
and photon wavefunctions~\cite{polar1,polar2,cc05}. Nevertheless, it has been proposed
in Refs.~\cite{berloff1,kbb,review_nb} that a single cGL equation
for the macroscopically occupied polariton state can also be used in a 
way consistent with experimental observations~\cite{nbb_nature}. 
%
A similar approach was followed in Ref.~\cite{magnons} where a BEC of
magnon quasi-particles, incorporating a source term rather than an 
amplification of the field, was shown to be phenomenologically described 
by a system two nonlinearly coupled cGL-type equations.
%
In the context of the single cGL model for the polaritons, 
there exists a localized (pumping) region
of gain and a nonlinear saturating loss term, in addition to all the
standard terms (quantum pressure, external parabolic trapping and
repulsive interatomic interaction) that one encounters
in atomic BECs~\cite{emergent}. Furthermore, it should be pointed out
that the prototypical setting where experiments have been
conducted is two-dimensional in nature. Yet, highly anisotropic traps
(similar to what has been done in atomic BECs~\cite{emergent}) can
be envisioned which reduce the effective dynamics to a quasi one-dimensional (1D)
setting~\cite{1d_polaritons1,1d_polaritons2,1d_polaritons3,1d_polaritons4,1d_polaritons5,1d_polaritons6}.
Moreover, recent experimental
advances have enabled the use of thin microwires in order to
guide the condensates along the direction of the wire~\cite{Amo:10}.
In this setting, the recent analysis of Ref.~\cite{augusto} presented a
number of striking characteristics due to the interplay of gain
and loss terms with the standard ones of atomic BECs. Prominent examples
included the destabilization of the nodeless state of the system and
the creation of stability inversions (where states with nodes would
be more robust), as well as the existence of bubble-like and sawtooth-like
solutions in the system.

A very interesting research direction in the physics of atomic and polariton BECs
concerns the dynamics of the condensates
in a double-well potential. The latter can be created in atomic
BEC experiments through the combination of a parabolic trap and a periodic
(so-called optical lattice) potential generated through the interference
of laser beams illuminating the BEC~\cite{Morsch}.
Relevant experiments
in atomic BECs~\cite{markus1,markus2}
have paved the way towards the exploration of numerous features such
as tunneling and Josephson oscillations for small numbers of atoms in the
condensate, and macroscopic quantum self-trapped states, as well as
symmetry-breaking effects for large atom
numbers. On the other hand, double-well potentials can also be created in
polariton BEC experiments in microcavities by applying stress \cite{balili,balapl}, by
employing photolithographic techniques \cite{1d_polaritons1,1d_polaritons2}, or
allowing natural formation during the sample growth \cite{lagdw}. Importantly, the latter
technique was used for the study of a ``polariton Josephson junction'' \cite{lagdw}, in
the spirit of earlier studies on ``bosonic Josephson junctions'' \cite{rudi} in the context of atomic BECs.
Importantly, a large volume of theoretical studies has accompanied
these developments, first in the context of atomic BECs, through investigations
related to finite-mode reductions and
symmetry-breaking
bifurcations~\cite{smerzi,kiv2,mahmud,bam,Bergeman_2mode,infeld,todd,TKF06},
quantum effects~\cite{carr}, and nonlinear variants of the double-well
potential \cite{pseudo}, and more recently in the context of polariton
condensates, especially as concerns Josephson oscillations
therein \cite{polardw}.
It should be mentioned in passing that similar
(spontaneous symmetry breaking) effects have been monitored in the realm
of nonlinear optics:
in this context, formation of asymmetric
states in dual-core fibers \cite{fibers}, self-guided laser beams in Kerr
media \cite{HaeltermannPRL02}, and optically-induced dual-core waveguiding
structures in photorefractive crystals \cite{zhigang} have been reported.

It is the aim of the present work to combine
these two themes, namely the focus on the exciton-polariton
BEC with pumping and loss and the fundamental interest in the
understanding of double-well trapping potentials in a spirit similar
to the proposal of Ref.~\cite{polar1}.
In particular, we will consider the single-component
model of Refs.~\cite{berloff1,kbb,review_nb} combined with a double-well
potential in a quasi-1D (e.g., microwire) setting.
We will attempt a systematic (Galerkin) finite-dimensional
reduction of the system
via projection to the two principal eigenstates of the potential,
and will derive a damped-driven system of ordinary differential
equations (ODEs) that have been shown in the Hamiltonian case to capture
the essence of the statics~\cite{kirr} and dynamics~\cite{marzuola}
of double-well potentials. We will then examine the bifurcation
structure of the resulting ODEs and compare it to that of the original
partial differential equation (PDE) model.
This already provides us with a number of interesting
features that distinguish this system from its Hamiltonian analog.
For instance, in the case of attractive interatomic interactions
(which is studied together with that of repulsive interactions)
the relevant symmetry-breaking pitchfork bifurcation is subcritical
instead of supercritical as in the Hamiltonian case. Furthermore,
both branches that emerge from the pitchfork bifurcations, the stable
asymmetric one and the (now) unstable ``parent'' branch, both appear
to become destabilized in this polariton BEC setting for slightly
larger nonlinearities, posing the natural question of what is the
stable dynamics for larger values of the nonlinearity. These questions
will in part be addressed via direct numerical simulations.

Our presentation will be structured as follows.
First, in Section II, we will present the model
and its theoretical study via the Galerkin analysis.
In Section III, we will
study the model numerically and
compare the results of
the numerical bifurcation analysis with the prediction of the Galerkin
approximation.
We will also complement these results with
direct numerical simulations of the original model.
Finally, in section IV, we summarize our results and present our conclusions.

\section{Model setup and analytical predictions}

In our analysis below, we adopt the model of Refs.~\cite{berloff1,kbb,review_nb}.
It has been argued in these works that the original exciton-polariton
system given by a set of two coupled equations can be effectively reduced
to a single cGL equation with a nonlinear saturating loss term.
This reduction can be used when the reservoir mean-field potential is negligible 
and the spot size is large compared with the condensate size (i.e., if we can
consider that the spot width is the same of the spatial extent of the system).
In particular, the amplification of the existing field introduces a gain and
hence acts as a generator of polaritons. Then the loss term saturates this
gain beyond a certain threshold. These two terms are analogous to the pumping
of polaritons from the excitons and to the natural decay of the polaritons.
%
This reduced model can be expressed in dimensionless form as follows:
\begin{equation}
i\partial_{t}u=-\partial_{x}^{2}u+s|u|^{2}u+V(x)u-\mu u+i\left[\chi(x)-\sigma|u|^{2}\right]u.
\label{eq:gpeq1}
\end{equation}
The above model is actually a complex Ginzburg-Landau equation \cite{aranson} for the complex order parameter
$u(x,t)$, which is assumed to evolve in the presence of
the effectively-1D double-well potential $V(x)$. Equation~(\ref{eq:gpeq1}) can
be applied
to both the contexts of atomic and polariton BECs: in the first case, the two last terms in the right-hand side of
Eq.~(\ref{eq:gpeq1}) are absent, and the model ---known as the Gross-Pitaevskii equation \cite{emergent}--- describes
the evolution of the macroscopic wavefunction for the cold atoms and $\mu$ is the chemical potential; in the second
case, $u(x,t)$ denotes the polariton wavefunction, and the last two terms in the right-hand side are included in the model.
More specifically, in the context of polariton condensates, Eq.~(\ref{eq:gpeq1}) incorporates
(a) the spatially dependent gain term of the form
\begin{equation}
\chi(x)=\alpha\Theta(x_m-|x|),
\end{equation}
where $\Theta$ is the step function generating a symmetric pumping spot
of ``radius'' $x_m$ and strength $\alpha$ for the gain, and (b)
a nonlinear saturation loss term,
characterized by its strength $\sigma$.
As concerns the parameter $s \equiv \pm 1$, it sets the type of nonlinearity (i.e., the type of interactions between
atoms or polaritons):
for $s=+1$ the nonlinearity is defocusing (i.e., the interactions are repulsive), while for $s=-1$ the nonlinearity is
focusing (i.e., the interactions are attractive). In the context of atomic BECs, the value of $s$ depends on the atom species
(e.g., $s=+1$ for $^{87}$Rb or $^{23}$Na, while $s=-1$ for $^7$Li or $^{85}$Rb atoms). On the other hand, in the context of polariton condensates, the sign of the effective mass of polaritons [i.e., the sign of the first term in the right-hand side
of Eq.~(\ref{eq:gpeq1})] may become either positive or negative, depending on the values of transverse momentum:
in fact, the transition from positive to negative mass is associated with the inflection point of the
energy-momentum diagram \cite{dvs}.
%
Here, we will consider both cases of $s = \pm 1$ to take into regard
that the effective polariton mass may be positive or negative, respectively.
We finally note that the relevant physical time and space scales,
as well as physically relevant parameter values associated with Eq.~(\ref{eq:gpeq1}), can be found in Ref.~\cite{berloff1}.

In what follows, we will use the Galerkin (few mode truncation) approach of Ref.~\cite{TKF06}. We start by considering
the corresponding linear eigenproblem which reads:
\begin{equation}
Hu \equiv -\partial_{x}^{2}u+V(x)u=\omega u,
\end{equation}
whose spectrum consists of a ground state, $u_0(x)$, and excited states, $u_i(x)$ (with $i \ge1$). Then, in the weakly
nonlinear regime, we consider a superposition of the two lowest linear eigenmodes,
\begin{equation}
u(x,t)=c_{0}(t)u_{0}(x)+c_{1}(t)u_{1}(x),
\label{ans}
\end{equation}
where $c_{0,1}(t)$ are unknown time-dependent complex prefactors;
obviously, the above ansatz is relevant
for values of the chemical potential $\mu$ such that
higher order modes can be safely ignored. Substituting this ansatz into Eq.~(\ref{eq:gpeq1})
we obtain:
\begin{equation}
i \left(\dot{c_{0}}u_{0}+\dot{c_{1}}u_{1}\right)
=(\omega_{0}-\mu)u_{0}c_{0}+(\omega_{1}-\mu)u_{1}c_{1}+s |u|^{2} \left(c_0 u_0
+c_1 u_1 \right)
+i\left[\chi(x)-\sigma|u|^{2}\right] \left(c_{0}u_{0}+c_{1} u_{1}\right),
\end{equation}
where the $|u|^2$ has not been expanded only for reasons of compactness but
should actually be thought as expanded according to Eq.~(\ref{ans}).
%
Next, projecting on $u_{0}$ and $u_{1}$ (i.e., multiplying
the above equation by
$u_{0}$ and $u_{1}$
and integrating over $x$), and using the orthogonality of the states $u_{i}$, we
respectively derive the following equations:
%
%
%
%
\begin{equation}
i \dot{c_{0}}  =(\omega_{0}-\mu+i\alpha_0)c_{0} +
(s-i\sigma)\left\{ A_{0}|c_{0}|^{2}c_{0}+\left(c_{0}^{2}c_{1}^{*}+
2|c_{0}|^{2}c_{1}\right)\Gamma_{0}
+\left(2|c_{1}|^{2}c_{0}+c_{1}^{2}c_{0}^{*}\right)B+|c_{1}|^{2}c_{1}\Gamma_{1}\right\},
\end{equation}
and
\begin{equation}
i \dot{c_{1}} =(\omega_{1}-\mu+i\alpha_1)c_{1} +
(s-i\sigma)\left\{ \Gamma_{0}|c_{0}|^{2}c_{0}+\left(c_{0}^{2}c_{1}^{*}+
2|c_{0}|^{2}c_{1}\right)B
+\left(2|c_{1}|^{2}c_{0}+c_{1}^{2}c_{0}^{*}\right)\Gamma_{1}+|c_{1}|^{2}c_{1}A_{1}\right\}.
\end{equation}
In the above equations, overdots denote time derivatives,
the involved constants (depending on the eigenbasis $\{u_i\}$) take the values
$A_0 = \int u_0^4 dx$, $A_1 = \int u_1^4 dx$, $B = \int u_0^2u_1^2 dx$,
$\Gamma_{0} = \int u_1 u_0^3 dx$, and
$\Gamma_{1} = \int u_0 u_1^3 dx$, while the effective gain coefficients read:
$\alpha_0 = \int \chi(x)u_0^2 dx$ and $\alpha_1 = \int \chi(x)u_1^2 dx$.
%
%
We now use amplitude and phase variables for the time-dependent prefactors,
i.e., $c_{i}=\rho_{i}e^{i\phi_{i}}$ (with the amplitudes $\rho_i$ and phases $\phi_i$
being real functions), to derive a set of four equations for the unknown functions $\rho_{0,1}$ and $\phi_{0,1}$.
Introducing the relative phase of the first two modes as $\varphi \equiv \phi_1-\phi_0$,
the above mentioned set of equations takes the following form:
\begin{equation}\label{eq:rho0}
\begin{split}
\dot{\rho}_{0}=&\alpha_0\rho_{0}-\sigma\left(A_{0}\rho_{0}^{3}+2B\rho_{1}^{2}\rho_{0}\right)
  +s\left(\Gamma_{1}\rho_{1}^{3}+\Gamma_{0}\rho_{0}^{2}\rho_{1}\right)\sin\varphi\\
&+sB\rho_{1}^{2}\rho_{0}\sin2\varphi
  -\sigma\left(\Gamma_{1}\rho_{1}^{3}+3\Gamma_{0}\rho_{0}^{2}\rho_{1}\right)\cos\varphi-\sigma B\rho_{1}^{2}\rho_{0}\cos2\varphi,
\end{split}
\end{equation}
%
%
%
%
\begin{equation}\label{eq:phi0}
\begin{split}
\dot{\phi_{0}}  =& -(\omega_{0}-\mu)-s\left(A_{0}\rho_{0}^{2}+2B\rho_{1}^{2}\right)
 -\sigma\left(\Gamma_{0}\rho_{0}\rho_{1}+\Gamma_{1}{\rho_{1}^{3}}/{\rho_{0}}\right)\sin\varphi\\
&-\sigma B\rho_{1}^{2}\sin2\varphi
 -s\left(3\Gamma_{0}\rho_{0}\rho_{1}+\Gamma_{1}{\rho_{1}^{3}}/{\rho_{0}}\right)\cos\varphi
 -sB\rho_{1}^{2}\cos2\varphi,
\end{split}
\end{equation}
%
%
%
\begin{equation}\label{eq:rho1}
\begin{split}
\dot{\rho}_{1} =&\alpha_1\rho_{1}-\sigma\left(A_{1}\rho_{1}^{3}+2B\rho_{0}^{2}\rho_{1}\right)
  -s\left(\Gamma_{0}\rho_{0}^{3}+\Gamma_{1}\rho_{1}^{2}\rho_{0}\right)\sin\varphi\\
&-sB\rho_{0}^{2}\rho_{1}\sin2\varphi
 -\sigma\left(\Gamma_{0}\rho_{0}^{3}+3\Gamma_{1}\rho_{1}^{2}\rho_{0}\right)\cos\varphi-\sigma B\rho_{0}^{2}\rho_{1}\cos2\varphi,
 \end{split}
\end{equation}
and
\begin{equation}\label{eq:phi1}
\begin{split}
\dot{\phi_{1}} =&-(\omega_{1}-\mu)-s\left(A_{1}\rho_{1}^{2}+2B\rho_{0}^{2}\right)
 +\sigma\left(\Gamma_{1}\rho_{1}\rho_{0}+\Gamma_{0}{\rho_{0}^{3}}/{\rho_{1}}\right)\sin\varphi\\
&+\sigma B\rho_{0}^{2}\sin2\varphi
 -s\left(3\Gamma_{1}\rho_{1}\rho_{0}+\Gamma_{0}{\rho_{0}^{3}}/{\rho_{1}}\right)\cos\varphi-sB\rho_{0}^{2}\cos2\varphi.\end{split}
\end{equation}
Subtracting Eq.~(\ref{eq:phi0}) from Eq.~(\ref{eq:phi1}),
we can readily obtain an equation for $\varphi$, namely:
%
%
\begin{equation}
\begin{split}
\dot{\varphi}  =&-\Delta\omega-s\left(A_{1}\rho_{1}^{2}-A_{0}\rho_{0}^{2}\right)-sB\left[2+\cos2\varphi\right]\left(\rho_{0}^{2}-\rho_{1}^{2}\right)
 -s\frac{\cos\varphi}{\rho_{0}\rho_{1}}\left(\Gamma_{0}\rho_{0}^{2}(\rho_{0}^{2}-3\rho_{1}^{2})+\Gamma_{1}\rho_{1}^{2}(3\rho_{0}^{2}-\rho_{1}^{2})\right)\\
 & +\sigma\frac{\sin\varphi}{\rho_{0}\rho_{1}}\left(\Gamma_{0}\rho_{0}^{2}(\rho_{0}^{2}+\rho_{1}^{2})+\Gamma_{1}\rho_{1}^{2}(\rho_{0}^{2}+\rho_{1}^{2})\right)+\sigma B\sin2\varphi(\rho_{0}^{2}+\rho_{1}^{2}),
\end{split}
\label{vf}
\end{equation}
where $\Delta\omega \equiv \omega_1-\omega_2$. This way, we have arrived to a system of three equations
[cf.~Eqs.~(\ref{eq:rho0}), (\ref{eq:rho1}) and (\ref{vf})] for the unknown functions $\rho_{0,1}$ and $\varphi$. These
equations are subject to an additional constraint stemming from the balance
condition $dN/dt=0$, where
$N\equiv \int_{-\infty}^{+\infty} |u|^2 dx$ is
the number of polaritons (mathematically the squared $L^{2}$ norm).
The evolution of the latter, can readily be found by multiplying Eq.~(\ref{eq:gpeq1})
by $u^{*}$, the complex conjugate of Eq.~(\ref{eq:gpeq1}) by $u$, and then adding and integrating the
resulting equations. It is straightforward to find that the condition for equilibrium is:
\begin{equation}
\int_{-\infty}^{+\infty} \left(\chi(x)-\sigma|u|^{2}\right)|u|^{2}dx=0.
\label{eq}
\end{equation}
%
%
Substituting Eq.~(\ref{ans}) into Eq.~(\ref{eq}), also using the polar decomposition for $c_i(t)$
[and assuming a definite ---even in our considerations---
parity for the function $\chi(x)$], we find that the balance condition (\ref{eq}) takes the form:
\begin{equation}
\begin{split}\left(\alpha_0\rho_{0}^{2}+\alpha_1\rho_{1}^{2}\right) & -\sigma\left(A_{0}\rho_{0}^{4}+\rho_{1}^{4}A_{1}+4\rho_{0}^{2}\rho_{1}^{2}B\right)
-4\sigma\left(\rho_{0}^{3}\rho_{1}\Gamma_{0}+\rho_{1}^{3}\rho_{0}\Gamma_{1}\right)\cos\varphi
-2\sigma\rho_{0}^{2}\rho_{1}^{2}B\cos2\varphi=0,
\end{split}
\end{equation}
%
%
which essentially fixes $\rho_{1}$ once $\rho_{0}$
and $\varphi$ are found and thus reducing the effective number of degrees of
freedom for our approximations to only two ($\rho_0$ and $\varphi$).
%
%
%

Below, we will consider the case of a
%
symmetric double-well potential, for which $\Gamma_{1}=\Gamma_{0}=0$.
%
In this case, Eqs.~(\ref{eq:rho0}), (\ref{eq:rho1}) and (\ref{vf}) are reduced to the following simpler form,
\begin{equation}
\begin{split}\dot{\rho}_{0} & =\alpha_0 \rho_{0}-\sigma\left(A_{0}\rho_{0}^{3}+2B\rho_{1}^{2}\rho_{0}\right)+sB\rho_{1}^{2}\rho_{0}\sin2\varphi-\sigma B\rho_{1}^{2}\rho_{0}\cos2\varphi,
\end{split}
\label{eq:ss_rho0}
\end{equation}
\begin{equation}
\begin{split}\dot{\rho}_{1} & =\alpha_1 \rho_{1}-\sigma\left(A_{1}\rho_{1}^{3}+2B\rho_{0}^{2}\rho_{1}\right)-sB\rho_{0}^{2}\rho_{1}\sin2\varphi-\sigma B\rho_{0}^{2}\rho_{1}\cos2\varphi,
\end{split}
\label{eq:ss_rho1}
\end{equation}
\begin{equation}
\begin{split}\dot{\varphi} & =-\Delta\omega-s\left(A_{1}\rho_{1}^{2}-A_{0}\rho_{0}^{2}\right)-sB\left[2+\cos2\varphi\right]\left(\rho_{0}^{2}-\rho_{1}^{2}\right)+\sigma B\sin2\varphi(\rho_{0}^{2}+\rho_{1}^{2}),
\end{split}
\label{eq:ss_phi}\end{equation}
while the equilibrium condition is accordingly simplified as:
%
\begin{equation}
\begin{split}\left(\alpha_0\rho_{0}^{2}+\alpha_1\rho_{1}^{2}\right) & -\sigma\left(A_{0}\rho_{0}^{4}+\rho_{1}^{4}A_{1}+4\rho_{0}^{2}\rho_{1}^{2}B\right)
-2\sigma\rho_{0}^{2}\rho_{1}^{2}B\cos2\varphi=0.
\end{split}
\end{equation}

We can now turn to the study of stationary solutions
(i.e., $\dot\rho_0=\dot\rho_1=\dot\varphi=0$)
resulting from the Galerkin truncation analysis.
Particularly, from Eq.~(\ref{eq:ss_rho0}) we obtain two possible solutions:
%
%
%
\begin{equation}
\begin{cases}
i)\quad\rho_{0}=0,\\
ii)\quad\alpha_0-\sigma\left(A_{0}\rho_{0}^{2}+2B\rho_{1}^{2}\right)+sB\rho_{1}^{2}\sin2\varphi-\sigma B\rho_{1}^{2}\cos2\varphi & =0,
\end{cases}
\label{eq:ss_sol0}
\end{equation}
%
%
while from Eq.~(\ref{eq:ss_rho1}) we obtain:
\begin{equation}
\begin{cases}
i)\quad\rho_{1}=0\\
ii)\quad\alpha_1-\sigma\left(A_{1}\rho_{1}^{2}+2B\rho_{0}^{2}\right)-sB\rho_{0}^{2}\sin2\varphi-\sigma B\rho_{0}^{2}\cos2\varphi & =0.
\end{cases}
\label{eq:ss_sol1}
\end{equation}
Next,
multiplying the nontrivial equilibria of
Eq.~(\ref{eq:ss_sol0}) by $\rho_{0}^{2}$, the one from Eq.~(\ref{eq:ss_sol1})
by $\rho_{1}^{2}$, and adding the resulting equations, we obtain:
%
%
\begin{equation}
\cos2\varphi=\frac{(\alpha_0\rho_{0}^{2}+\alpha_1\rho_{1}^{2})-\sigma\left(A_{0}\rho_{0}^{4}+A_{1}\rho_{1}^{4}+4B\rho_{0}^{2}\rho_{1}^{2}\right)}{2\sigma B\rho_{0}^{2}\rho_{1}^{2}},
\label{eq:ss_sol3}
\end{equation}
while subtracting Eq.~(\ref{eq:ss_sol1}) from Eq.~(\ref{eq:ss_sol0}) yields:
\begin{equation}
\sigma\left(A_{1}\rho_{1}^{4}-A_{0}\rho_{0}^{4}+2B(\rho_{0}^{2}-\rho_{1}^{2})+
B(\rho_{0}^{2}-\rho_{1}^{2})\cos2\varphi\right)+sB(\rho_{0}^{2}+\rho_{1}^{2})\sin2\varphi
+(\alpha_0 - \alpha_1)=0.
\label{eq:ss_sol4}
\end{equation}
Combining now Eq.~(\ref{eq:ss_sol4}) with Eq.~(\ref{eq:ss_phi}) we finally
obtain the result:
%
\begin{equation}
(\rho_{0}^{2}+\rho_{1}^{2})\sin2\varphi=\frac{\sigma\Delta\omega -s(\alpha_0 - \alpha_1)}{B(\sigma^{2}+s^{2})}.
\label{sd}
\end{equation}

Let us now focus again on Eqs.~(\ref{eq:ss_rho0}) and (\ref{eq:ss_rho1}):
it is clear that if Eq.~(\ref{eq:ss_rho1}) is satisfied for
$\rho_{1}=0$
then $\rho_{0}^{2}=\frac{\alpha_0}{\sigma A_{0}}$,
and if Eq.~(\ref{eq:ss_rho0}) is satisfied with $\rho_{0}=0$
then $\rho_{1}^{2}=\frac{\alpha_1}{\sigma A_{1}}$.
Aside from these trivial symmetric and anti-symmetric solutions, past the
critical point for the symmetry breaking bifurcation, an asymmetric
solution is expected to exist which possesses non-vanishing $\rho_0$
and $\rho_1$ (as well as a non-zero relative phase between them), which can be
computed from Eq. (\ref{eq:ss_sol3}). It is anticipated that the presence
of loss and gain will not (generically) modify the nature of the bifurcations
in comparison to the Hamiltonian case
\cite{TKF06}. Namely,
an asymmetric solution will bifurcate from the symmetric one in
the focusing nonlinearity case of $s=-1$, due to a non-vanishing contribution of
the anti-symmetric part in the solution, while on the contrary,
an asymmetric mode will emanate from the anti-symmetric one in the
defocusing nonlinearity setting of $s=1$ (due to a symmetric contribution
within the solution). These results are detailed for a particular
case example potential in what follows and compared to full numerical
results.

\section{Numerical results}

\begin{figure}[htpb]
\begin{center}
    \includegraphics[width=8.5cm]{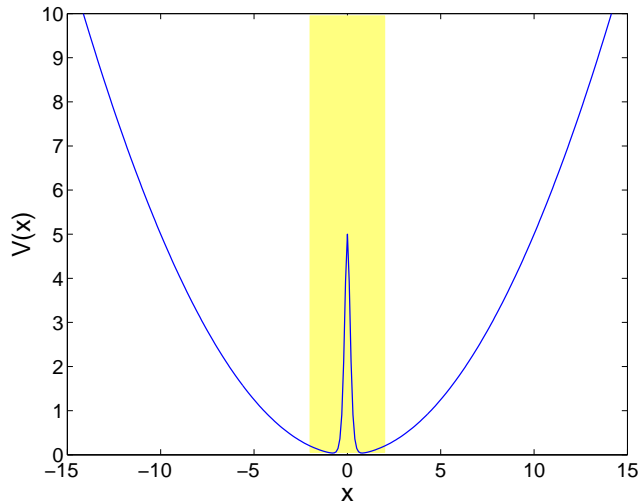}
\vspace{-0.3cm}
\caption{(Color online) The parabolic trapping potential and the localized barrier
creating the double-well potential configuration. The parameter values used
are: $\Omega^2=0.1$, $w=0.2$, and $V_0=5$; the shaded area
corresponds to the region where the pumping acts, i.e., $|x|<x_m=2$.}
\label{fig:potential}
\end{center}
\end{figure}

In our theoretical approximations, the double-well potential is
constructed by placing a localized barrier at the center of the parabolic trap
potential of strength $\Omega$. Particularly, the double-well potential
is assumed to be of the form:
%
\begin{equation}\label{eq:potential}
V(x)=\frac{1}{2}\Omega^{2}x^{2}+V_{0}\,\mathrm{sech}\left(\frac{x}{w}\right),
\end{equation}
where $w$ is the width of the barrier and $V_{0}$ its height.
The results presented below are for the potential parameters $\Omega^2=0.1$,
$V_{0}=5$, and $w=0.2$; we have checked that other parameter values lead to
quantitatively similar results .
For the gain we consider a strength $\alpha=0.2$ and a spot size of
$x_{m}=2.0$. The
damping parameter $\sigma$
is used to vary the number of atoms, $N$, in order to do the continuation.
For the above double-well potential, the values of the linear  eigen-energies
are $\omega_{0}=0.515729$ and $\omega_{1}=0.677697$. The
potential setting under consideration is
depicted in Fig.~\ref{fig:potential}.

\begin{figure}[htpb]
\begin{center}
\begin{tabular}{cc}
    \includegraphics[width=8.75cm]{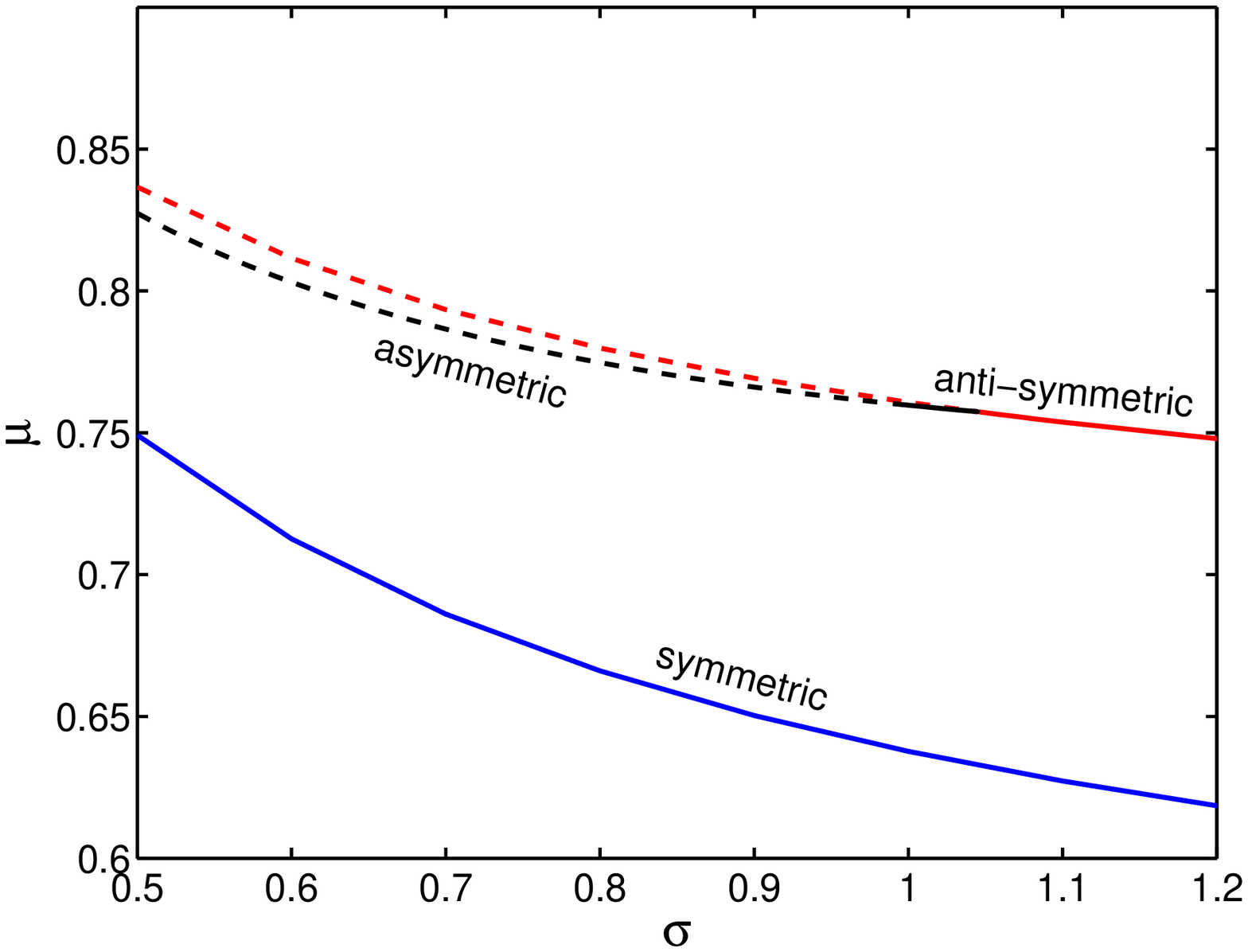} &
    \includegraphics[width=8.5cm]{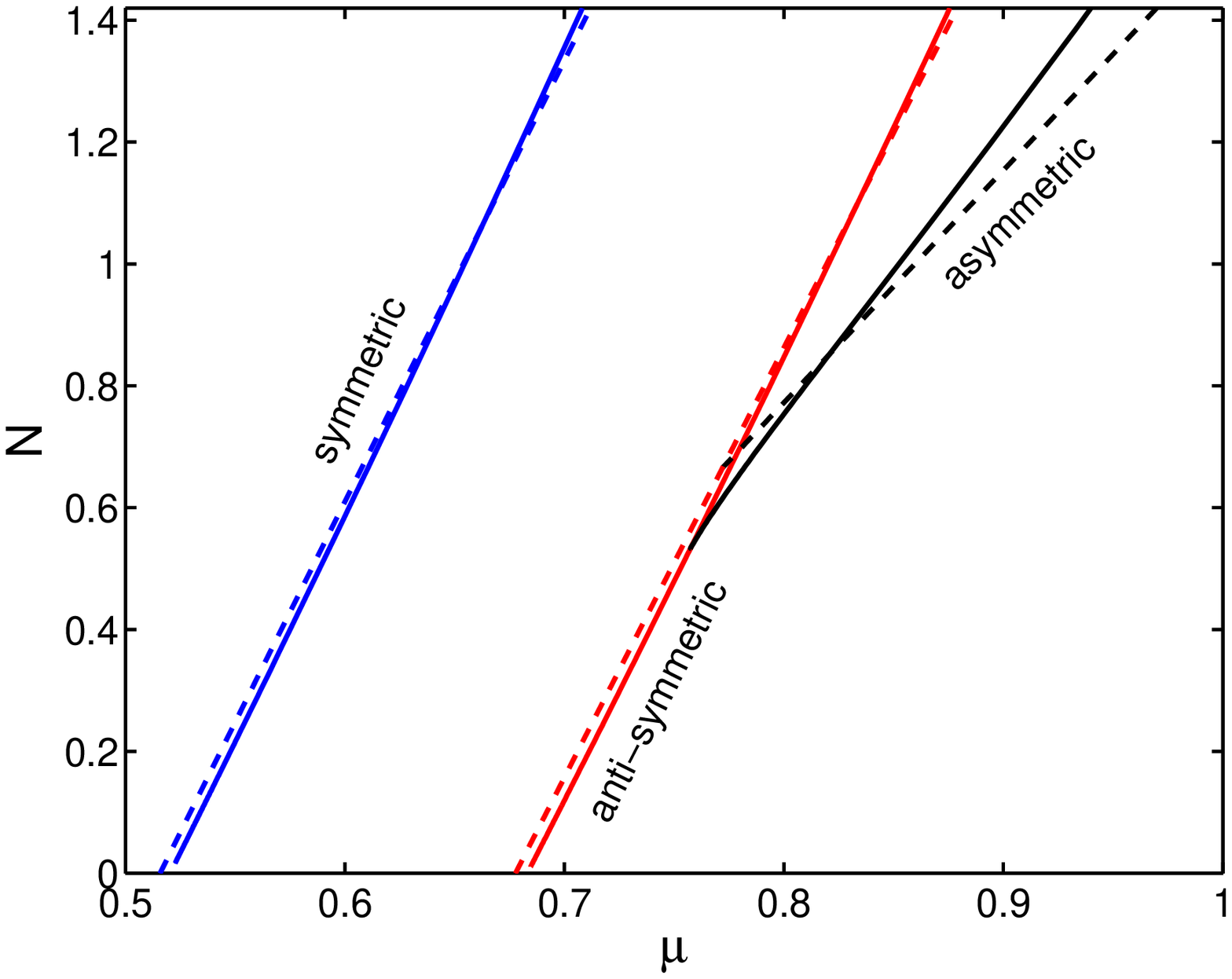}
\end{tabular}
\vspace{-0.3cm}
\caption{(Color online) 
Bifurcation diagrams for the symmetric, anti-symmetric and asymmetric
branches for defocusing (repulsive) nonlinearity ($s=1$).
Left: Dependence of the chemical potential
on the damping parameter.
Right: Dependence of the (normalized) number of polaritons
on the chemical potential. 
Unstable solutions are depicted by dashed lines on
the left panel. On the right solid lines display numerical
results obtained by a nonlinear (Newton-Raphson) solver of the
steady state equations of the model of Eq.~(\ref{eq:gpeq1}), while dashed lines
display analytical results of our Galerkin approach. The linear modes
are located at $\mu=0.5157$ and $\mu=0.6667$.}
\label{fig:chemrep}
\end{center}
\end{figure}

\begin{figure}[htpb]
\begin{center}
\begin{tabular}{cc}
    \includegraphics[width=8.2cm]{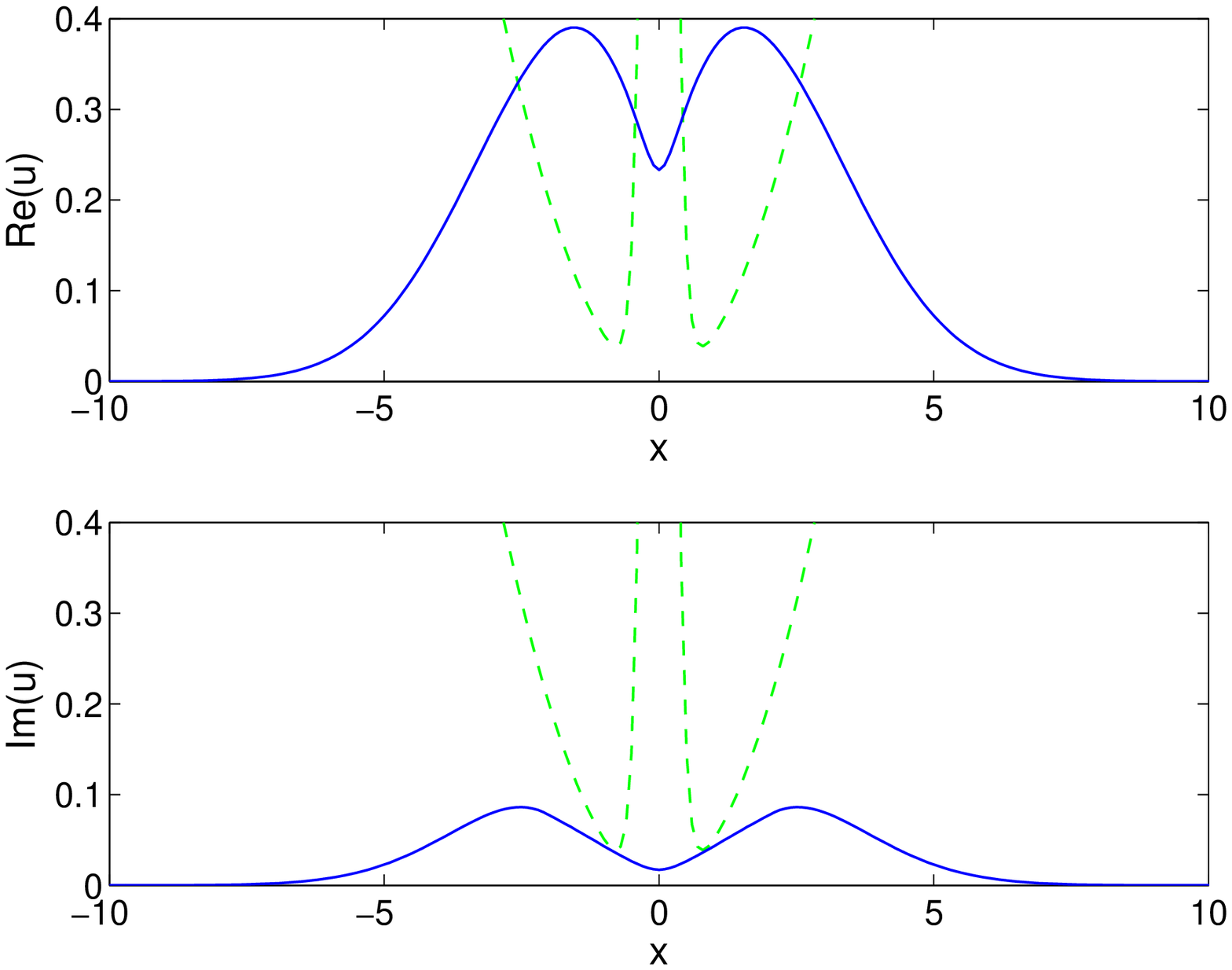} &~~
    \includegraphics[width=8.5cm]{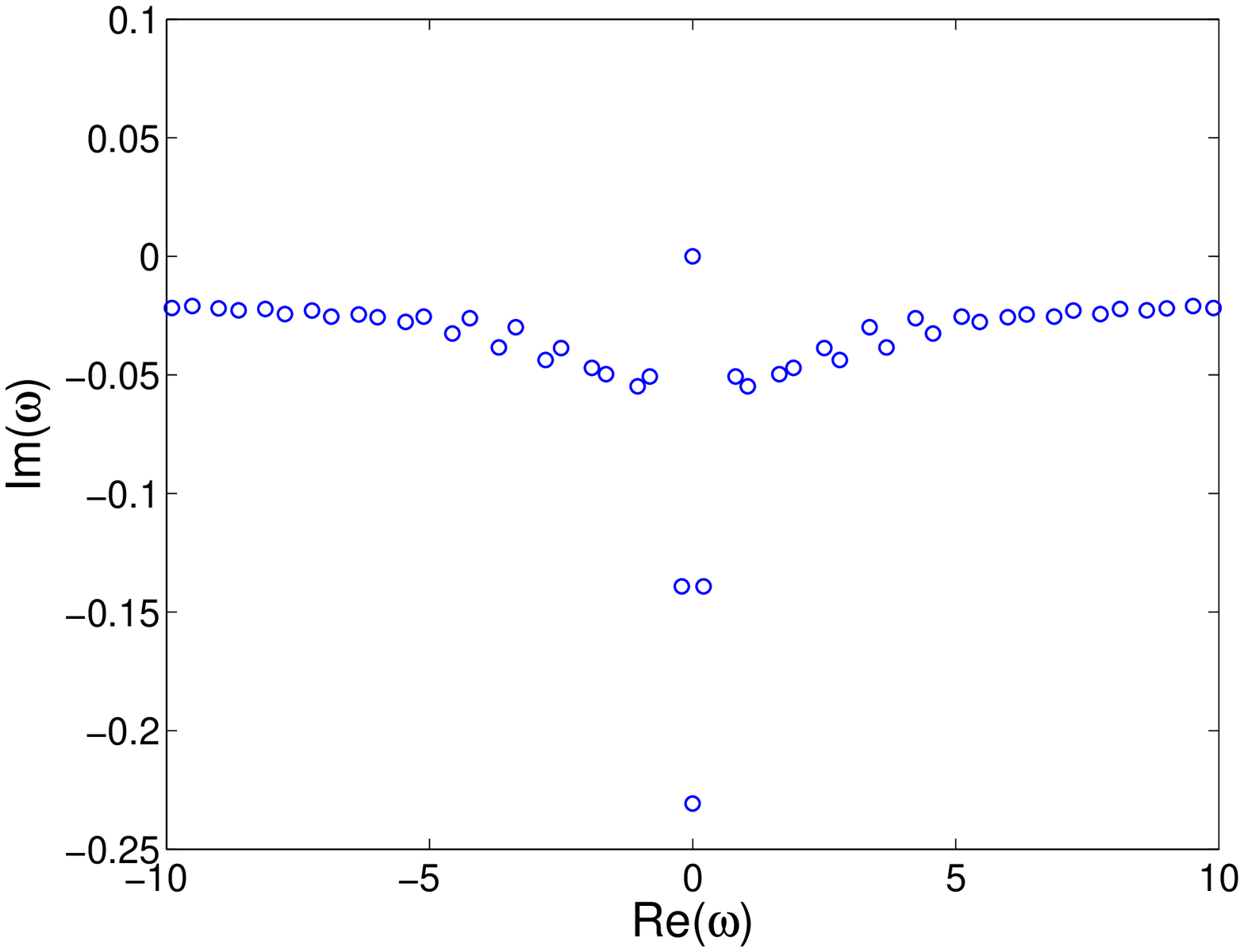} \\
    \includegraphics[width=8.2cm]{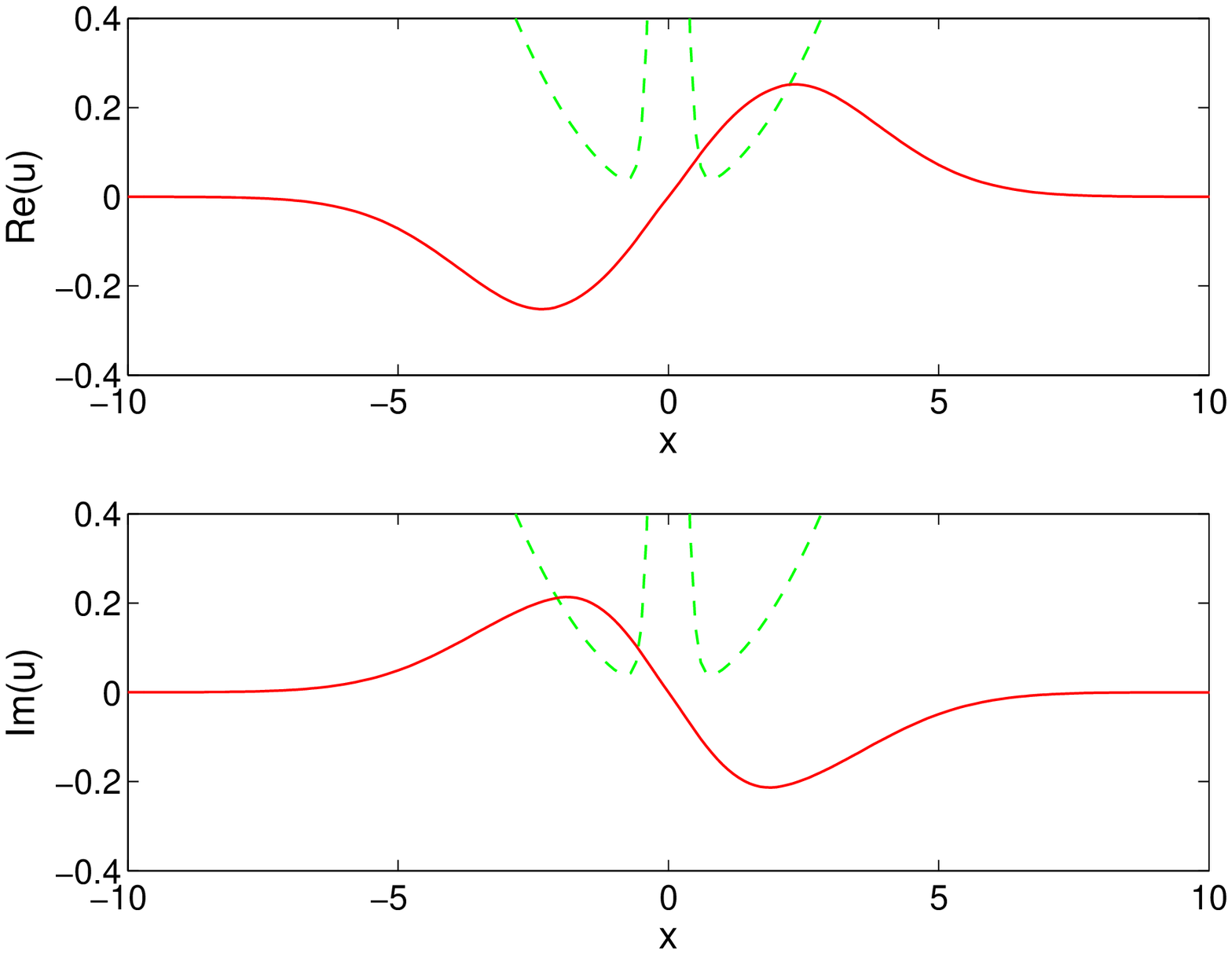} &~~
    \includegraphics[width=8.5cm]{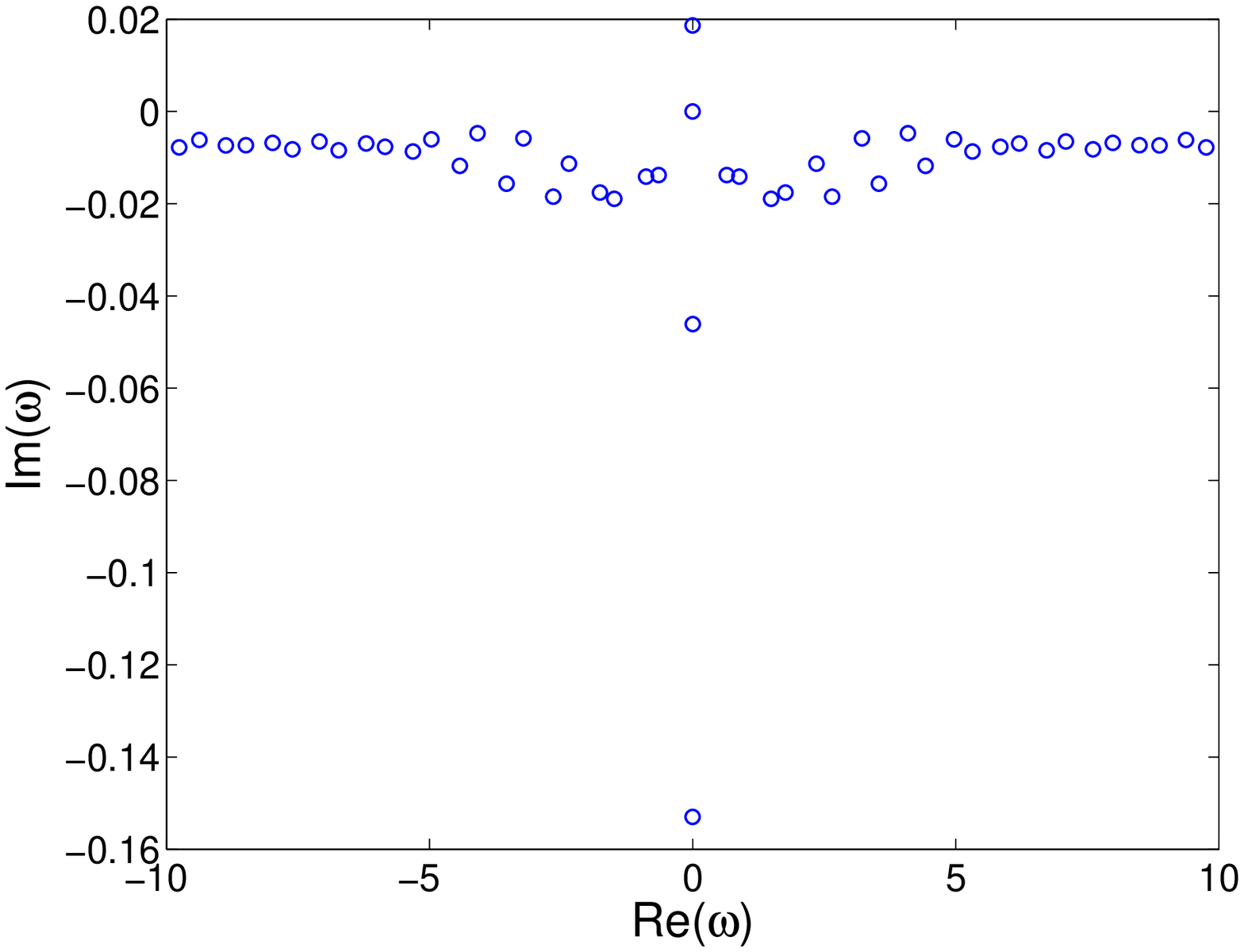} \\
\end{tabular}
\vspace{-0.3cm}
\caption{(Color online) (Left) Real and imaginary part of the wavefunction profile for a symmetric (top) and anti-symmetric (bottom) solution. (Right) Their corresponding stability eigenvalues. In all cases $\sigma=1$ and the
interactions are repulsive ($s=+1$).}
\label{fig:profrep1}
\end{center}
\end{figure}

\begin{figure}[htpb]
\begin{center}
\begin{tabular}{cc}
    \includegraphics[width=8.3cm]{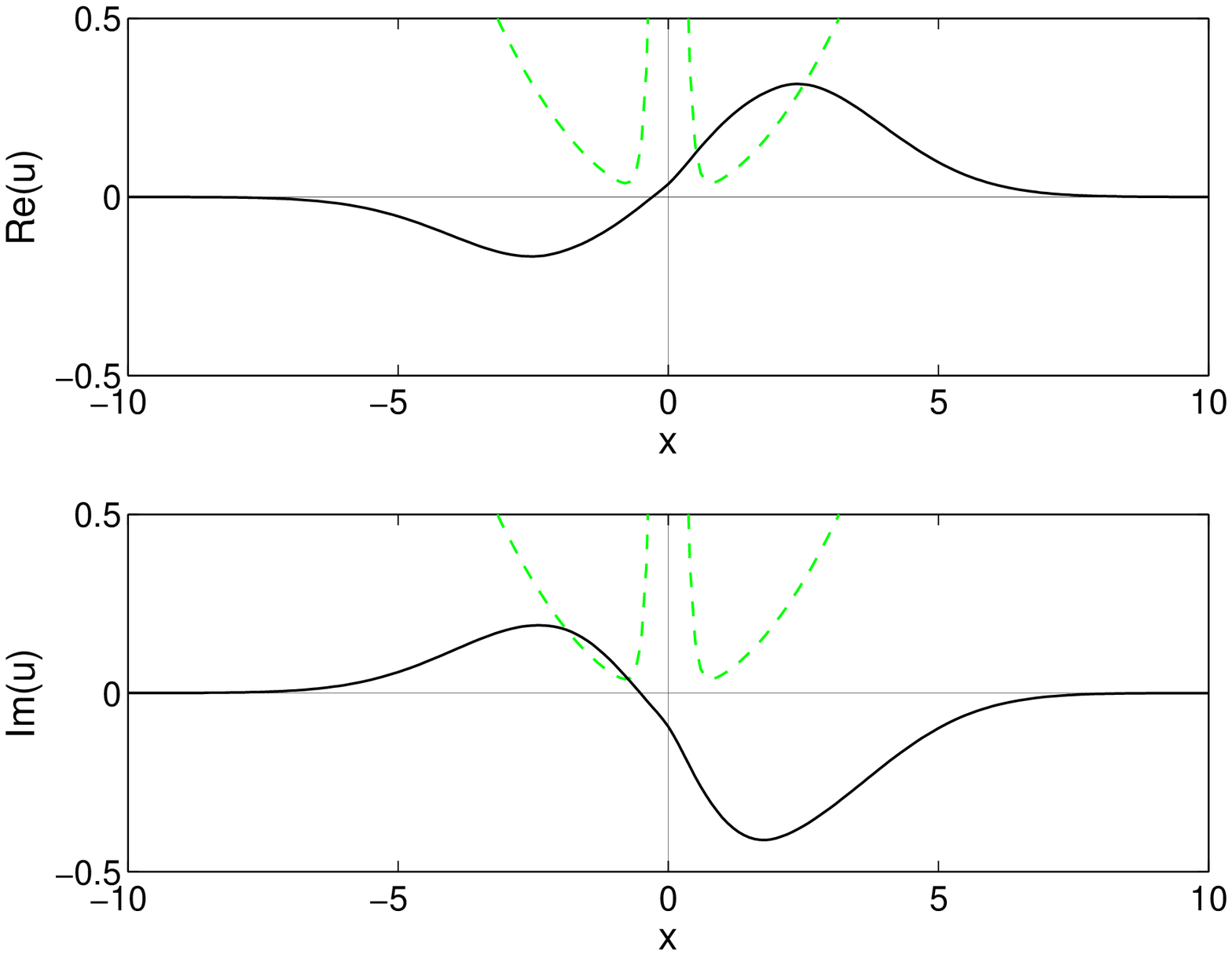} &~~
    \includegraphics[width=8.5cm]{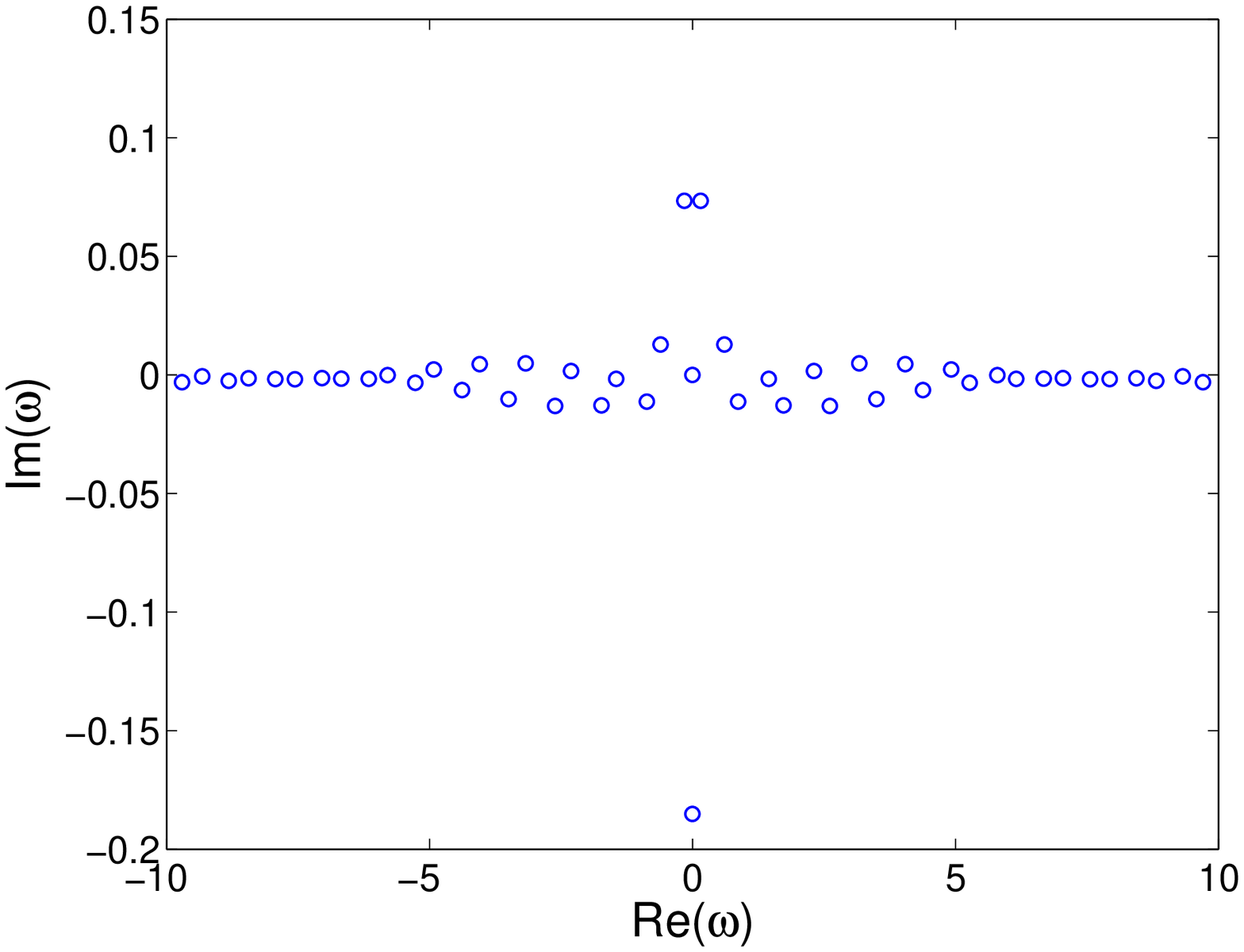} \\
\end{tabular}
\vspace{-0.3cm}
\caption{(Color online) (Left) Real and imaginary part of the wavefunction profile for an asymmetric solution with $\sigma=0.5$. (Right) Their corresponding stability
eigenvalues. All cases correspond to the repulsive interaction case ($s=+1$).}
\label{fig:profrep2}
\end{center}
\end{figure}

\begin{figure}[htpb]
\begin{center}
\begin{tabular}{cc}
    \includegraphics[width=8.5cm]{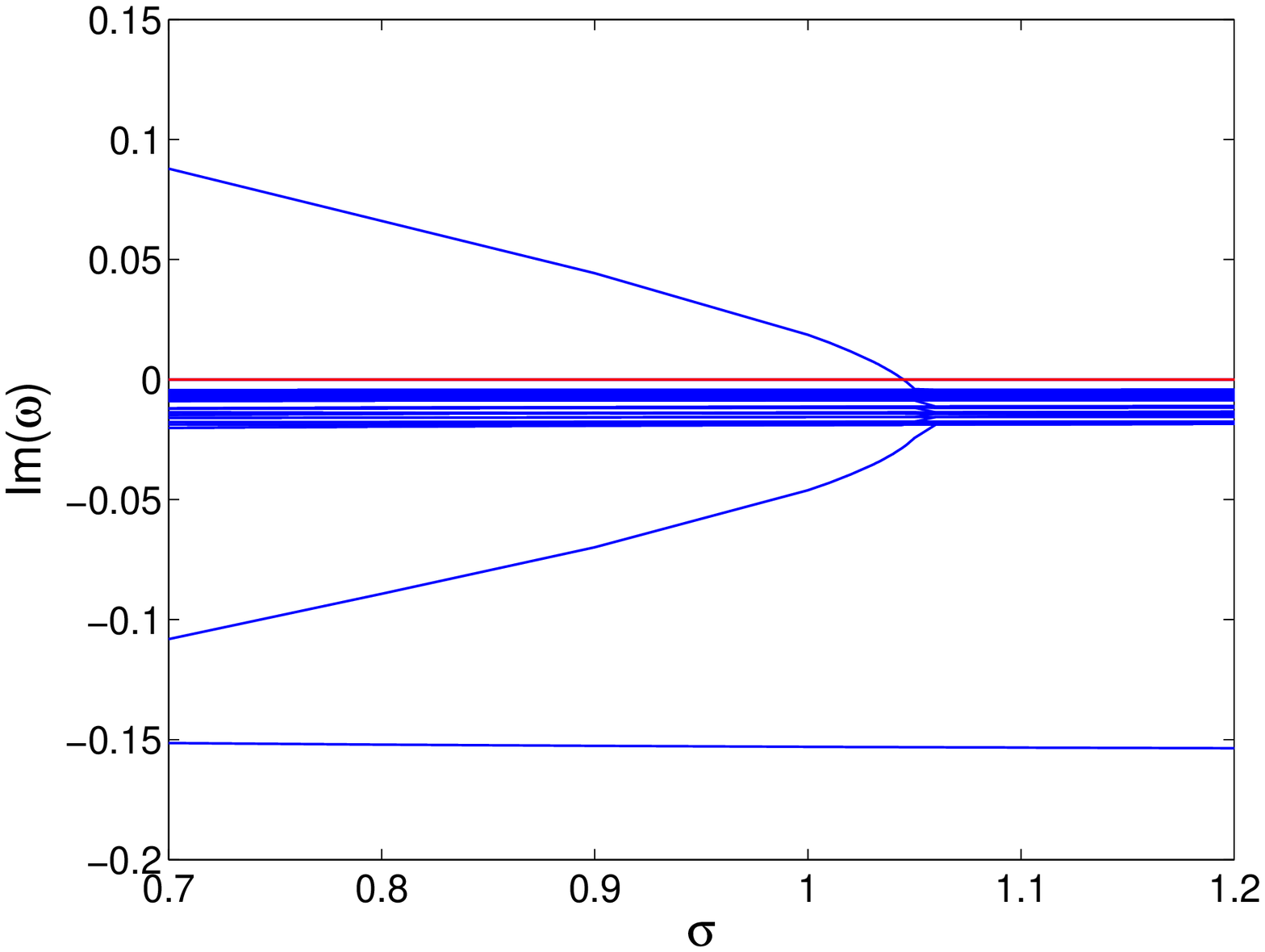} &
    \includegraphics[width=8.3cm]{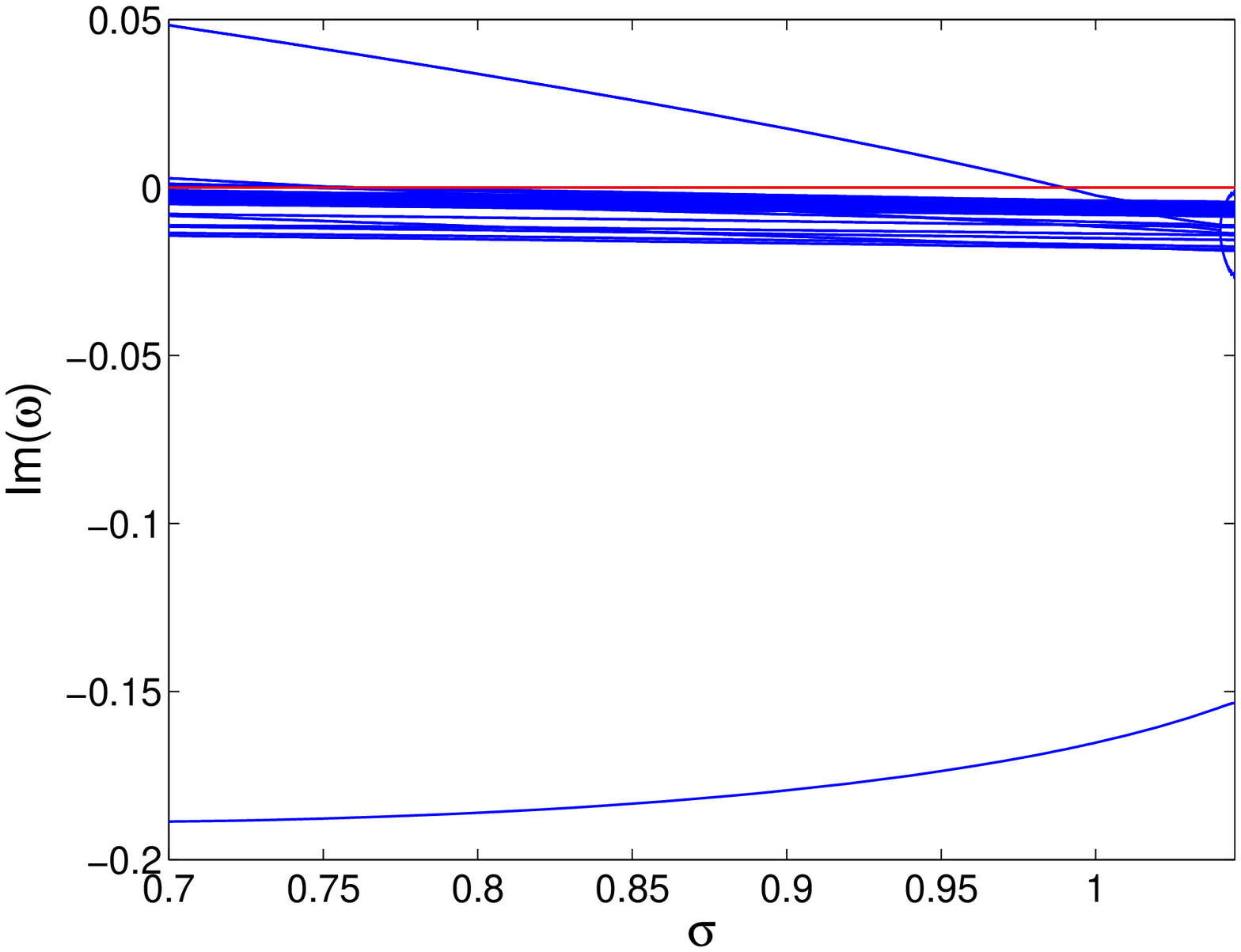} \\
\end{tabular}
\vspace{-0.3cm}
\caption{(Color online) Dependence of the imaginary part of the stability eigenvalues with respect to $\sigma$ for the anti-symmetric (left) and asymmetric
solutions (right) in the repulsive interaction case ($s=+1$).}
\label{fig:stabrep}
\end{center}
\end{figure}

\begin{figure}[htpb]
\begin{center}
\begin{tabular}{cc}
    \includegraphics[width=8.25cm,height=5.6cm]{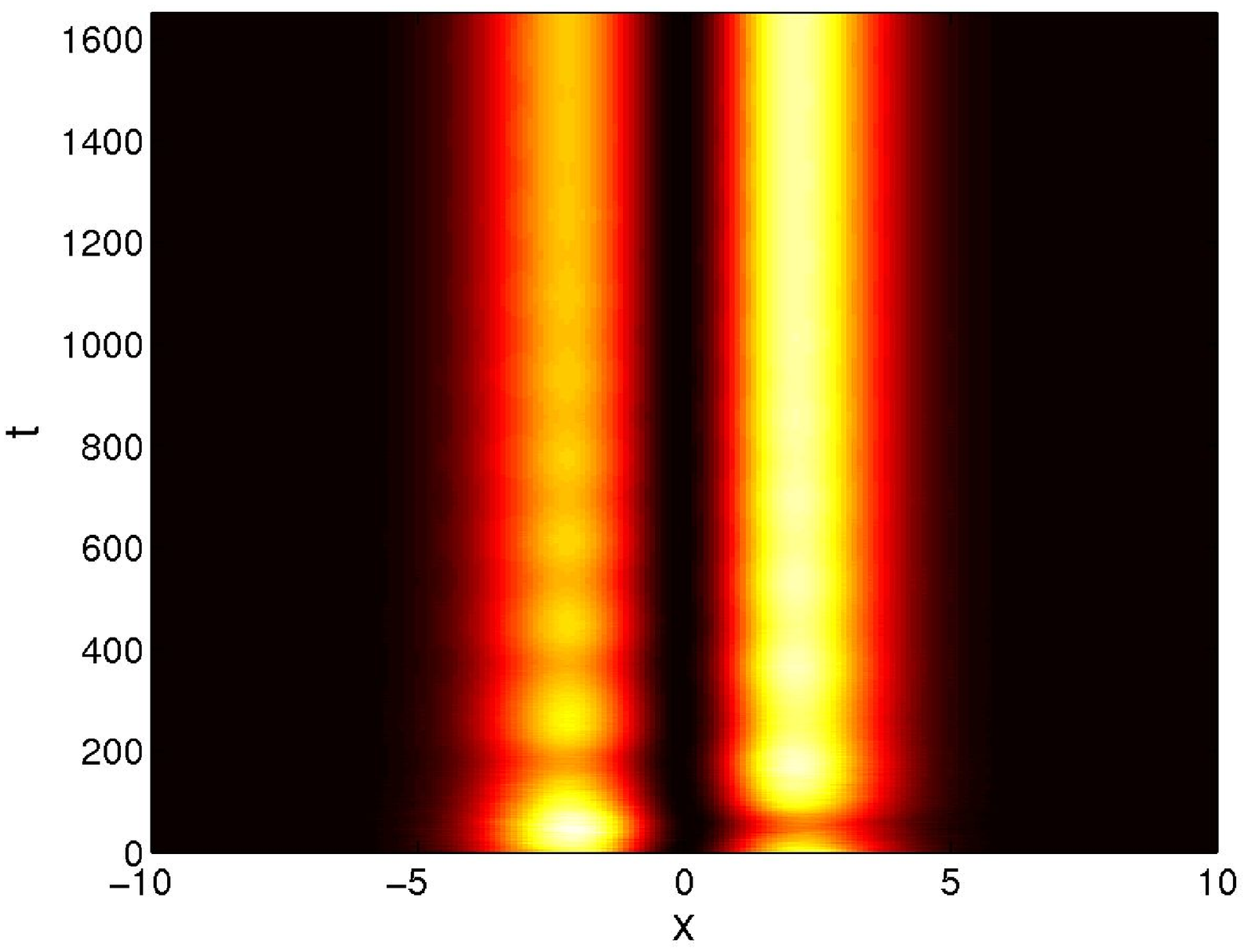} &
    \includegraphics[width=8.25cm,height=5.6cm]{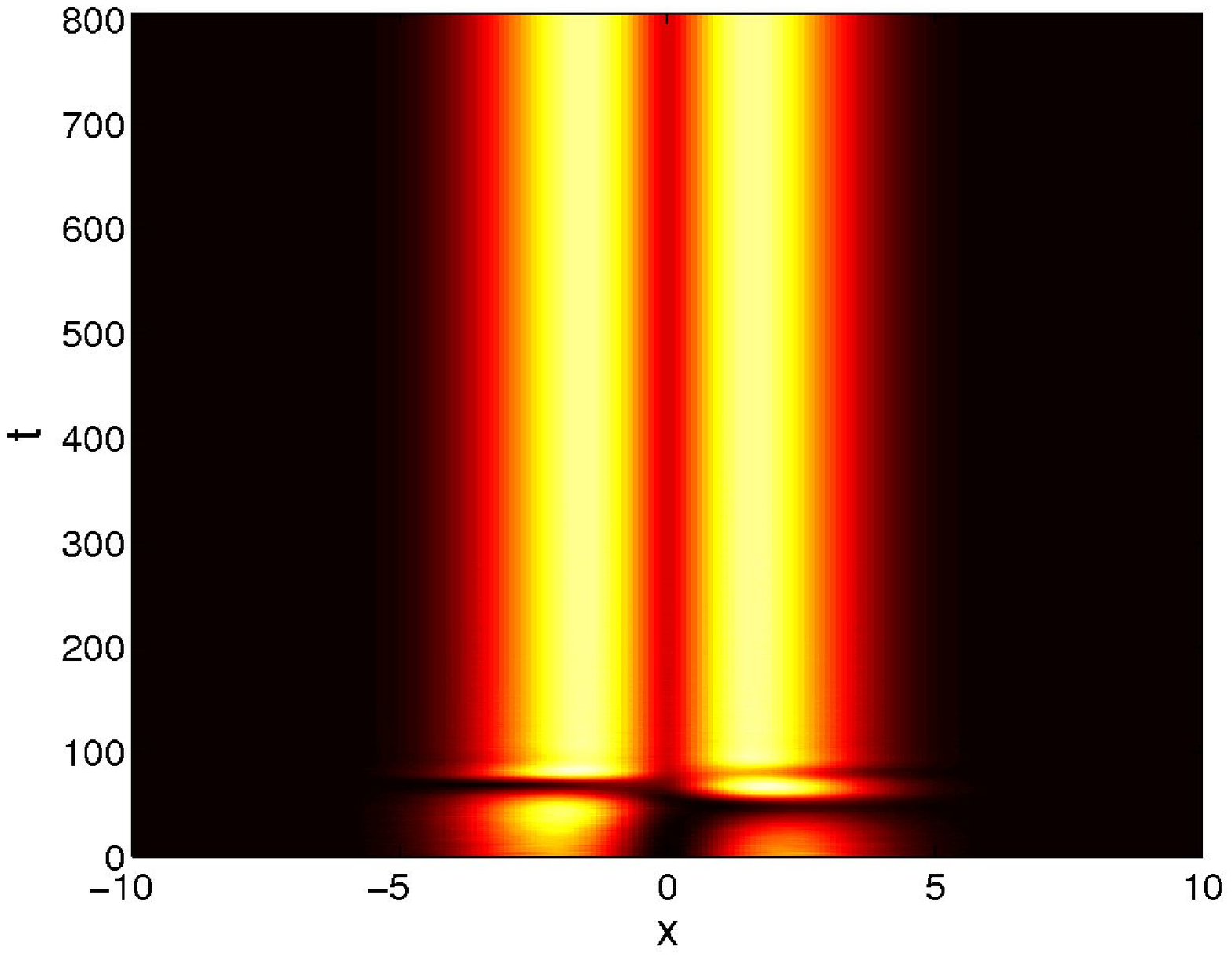} \\
\end{tabular}
    \includegraphics[width=17.2cm]{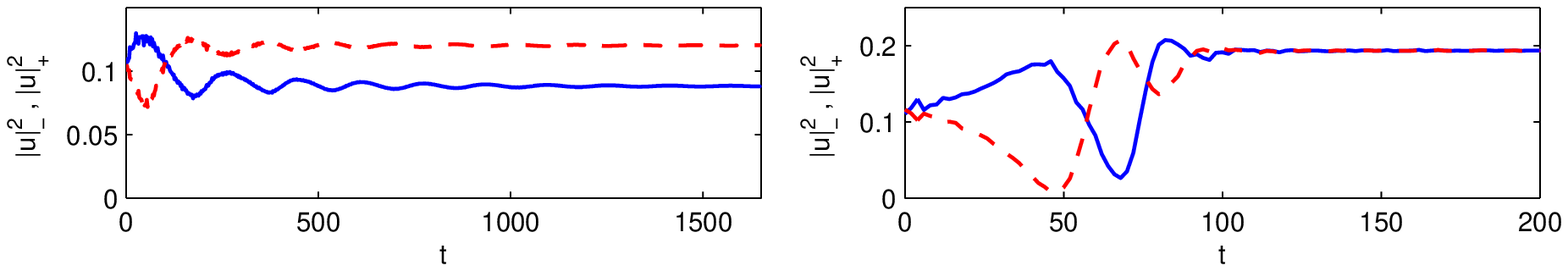}~
\vspace{-0.3cm}
\caption{(Color online) Top: Evolution of a perturbed anti-symmetric soliton
for $\sigma=1$ (left) and $\sigma=0.8$ (right) in the repulsive case
($s=+1$). The former case relaxes to the asymmetric stationary state,
while the latter to the symmetric ground state.
Bottom: Respective time
series for the density at the bottom of the left (solid blue line) and right
(dashed red line) wells.
}
\label{fig:dynrepanti}
\end{center}
\end{figure}

\begin{figure}[htpb]
\begin{center}
\begin{tabular}{cc}
    \includegraphics[width=8.25cm,height=5.6cm]{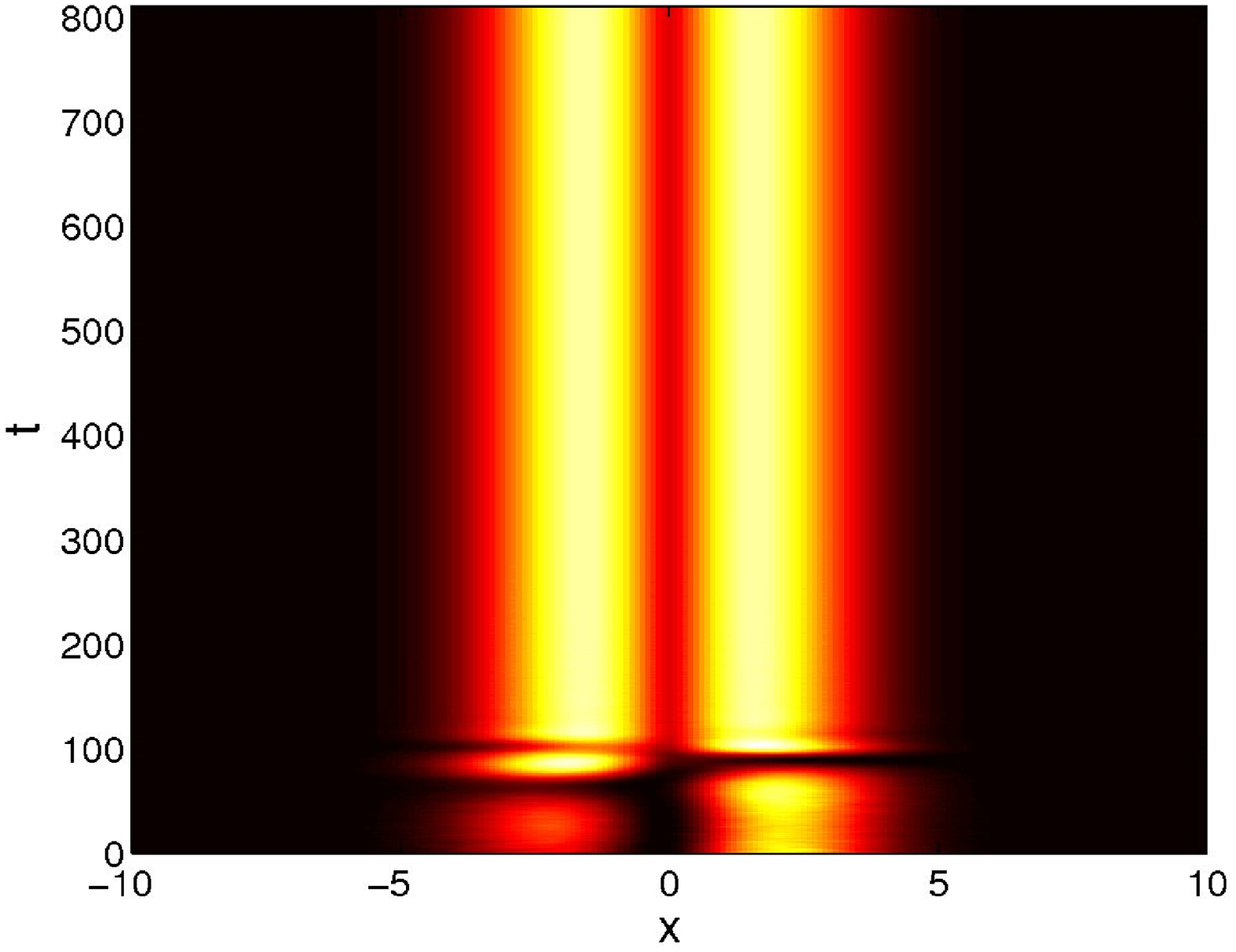} &
    \includegraphics[width=8.25cm,height=5.6cm]{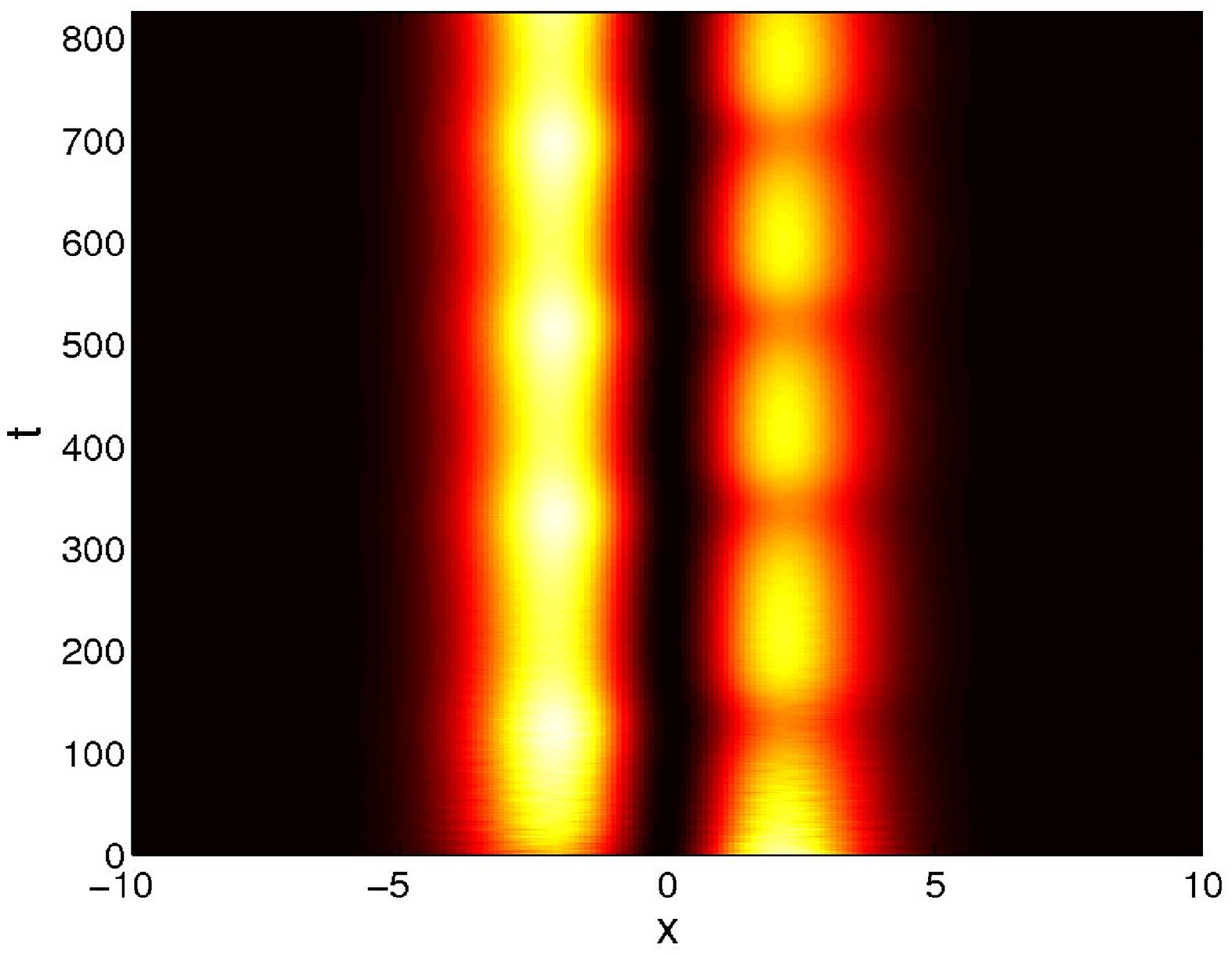} \\
\end{tabular}
    \includegraphics[width=17.2cm]{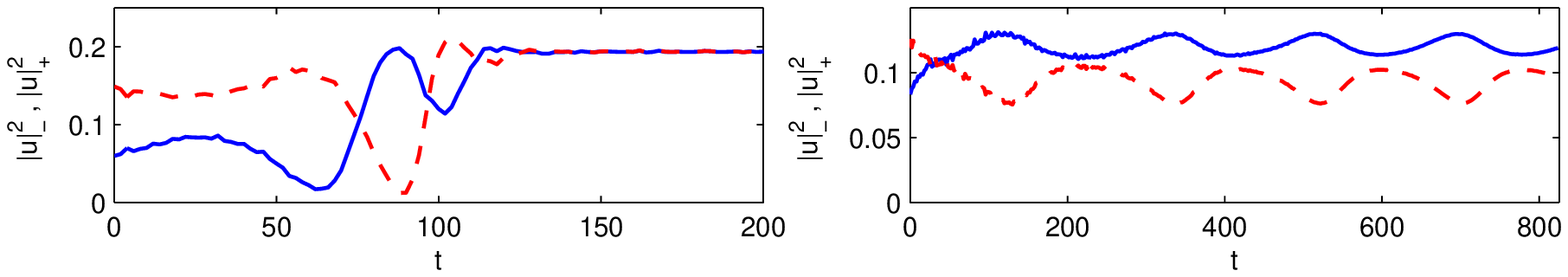}~~
\vspace{-0.3cm}
\caption{(Color online) Top: Evolution of a perturbed
asymmetric soliton with $\sigma=0.8$ (left) and $\sigma=0.98$ (right) in the repulsive case ($s=+1$). In the former case, perturbation leads to
the symmetric ground state attractor, while in the latter case,
it relaxes to a non-stationary (quasi-periodic) solution.
Bottom: Respective time
series for the density at the bottom of the left (solid blue line) and right
(dashed red line) wells.
}
\label{fig:dynrepasym}
\end{center}
\end{figure}

\begin{figure}[htpb]
\begin{center}
\begin{tabular}{cc}
    \includegraphics[width=8.7cm]{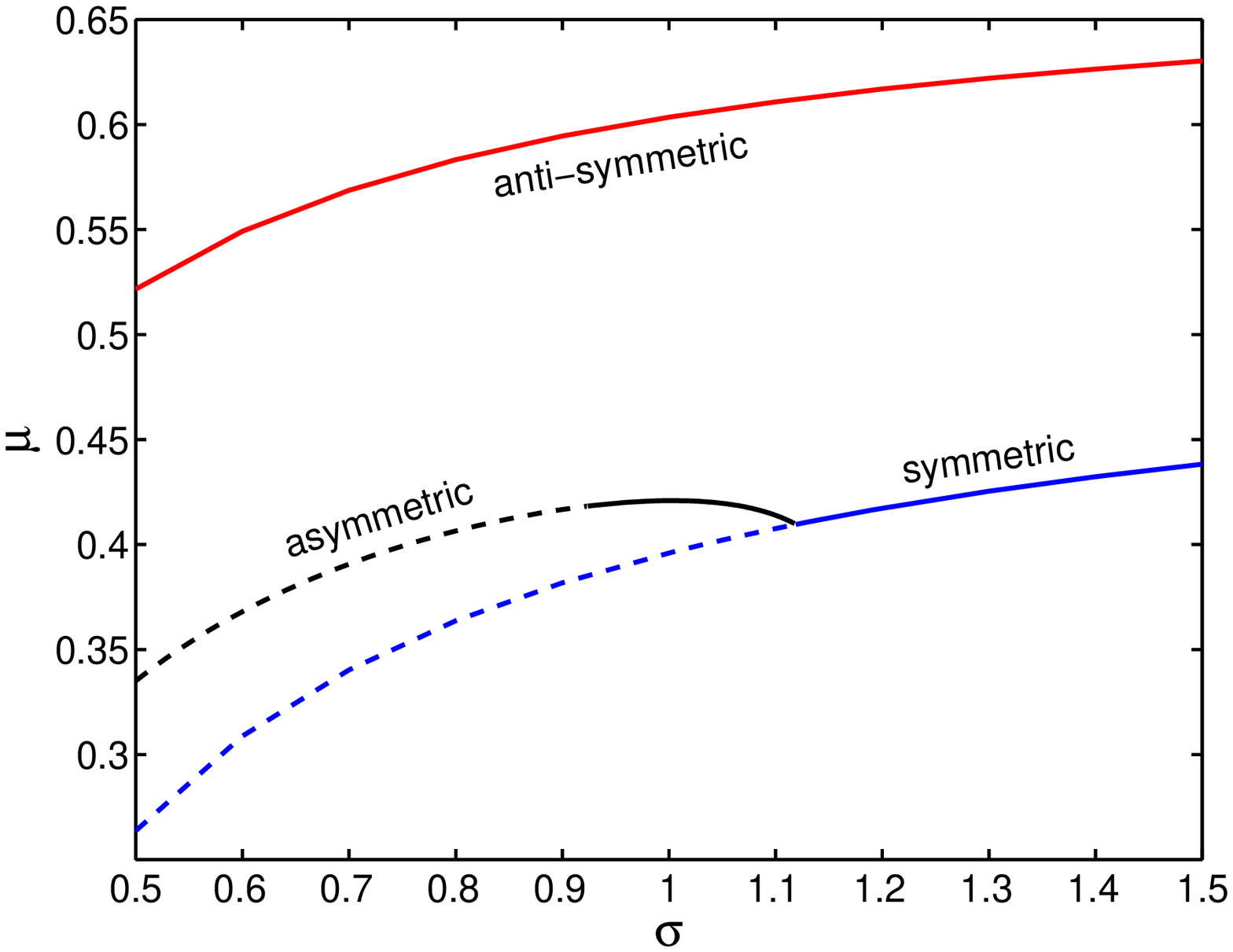} &
    \includegraphics[width=8.5cm]{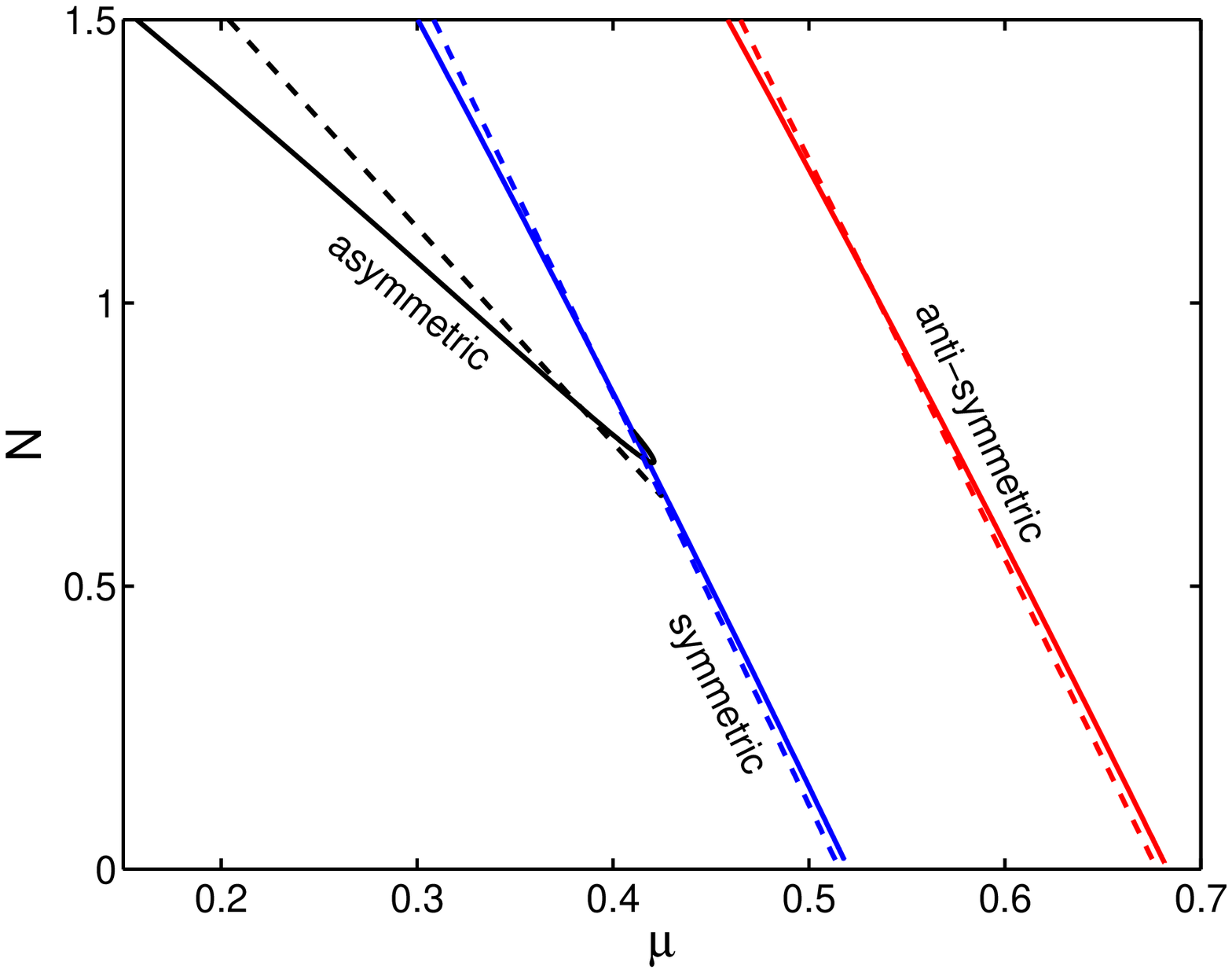}
\end{tabular}
\vspace{-0.3cm}
\caption{(Color online)
Bifurcation diagrams for the symmetric, anti-symmetric and asymmetric
branches for focusing (attractive) nonlinearity ($s=-1$).
Left: Dependence of the chemical potential
on the damping parameter.
Right: Dependence of the (normalized) number of polaritons
on the chemical potential.
Unstable solutions
are depicted by dashed lines on the left panel. On the right panel solid lines
display numerical (Newton-Raphson) results, while dashed lines
display analytical (Galerkin) results.}
\label{fig:chematt}
\end{center}
\end{figure}

\begin{figure}[htpb]
\begin{center}
\begin{tabular}{cc}
    \includegraphics[width=8.2cm]{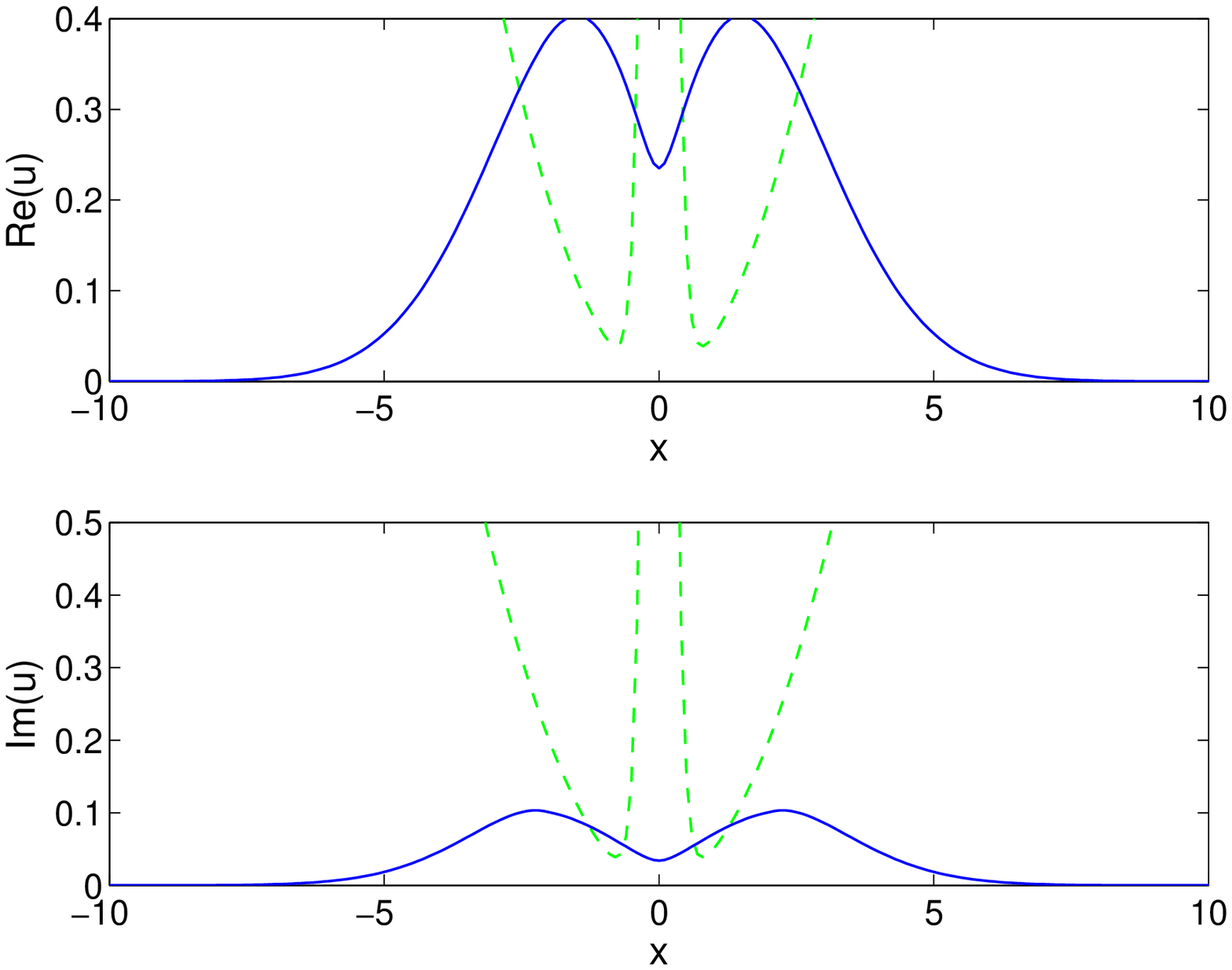} &~~
    \includegraphics[width=8.5cm]{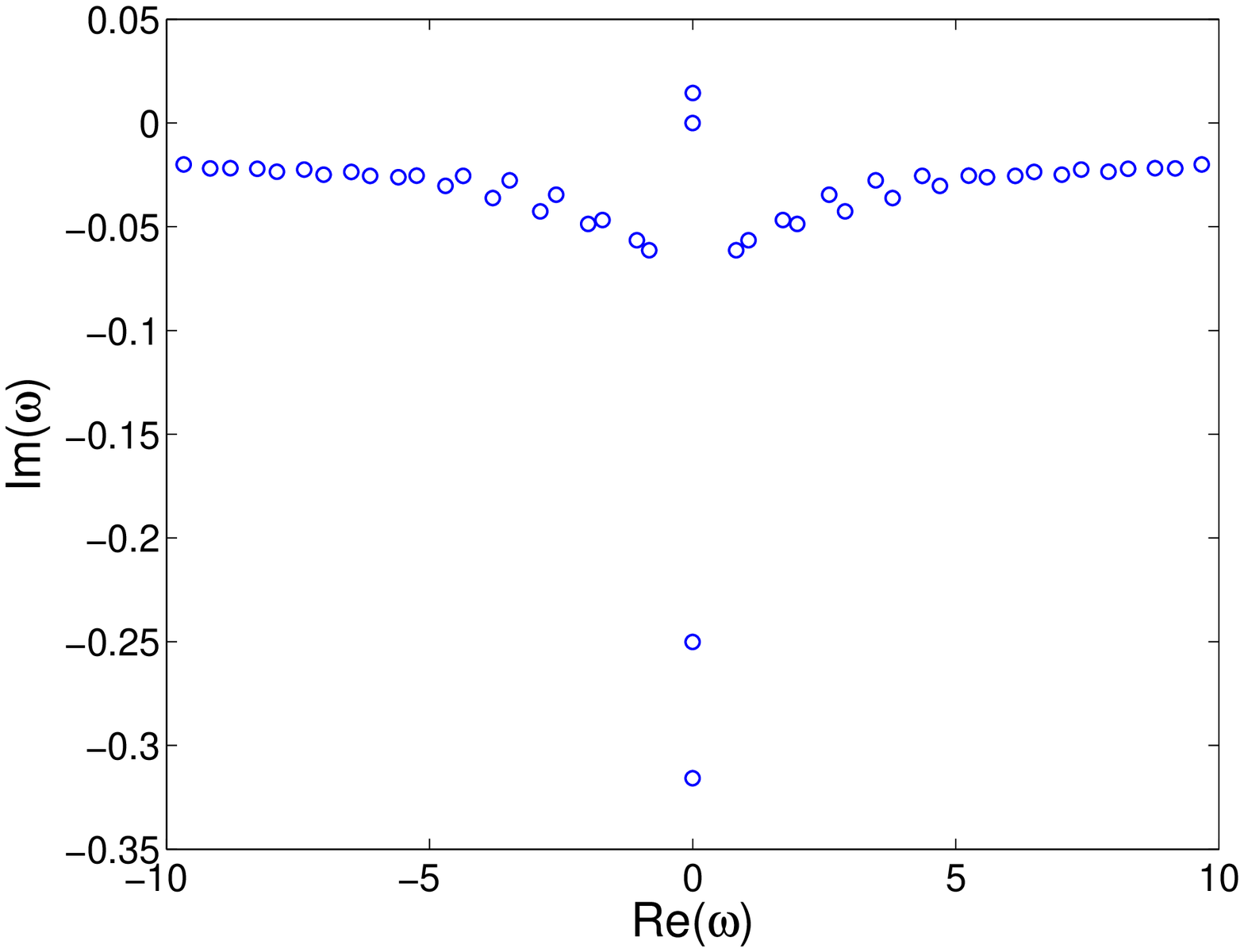} \\
    \includegraphics[width=8.2cm]{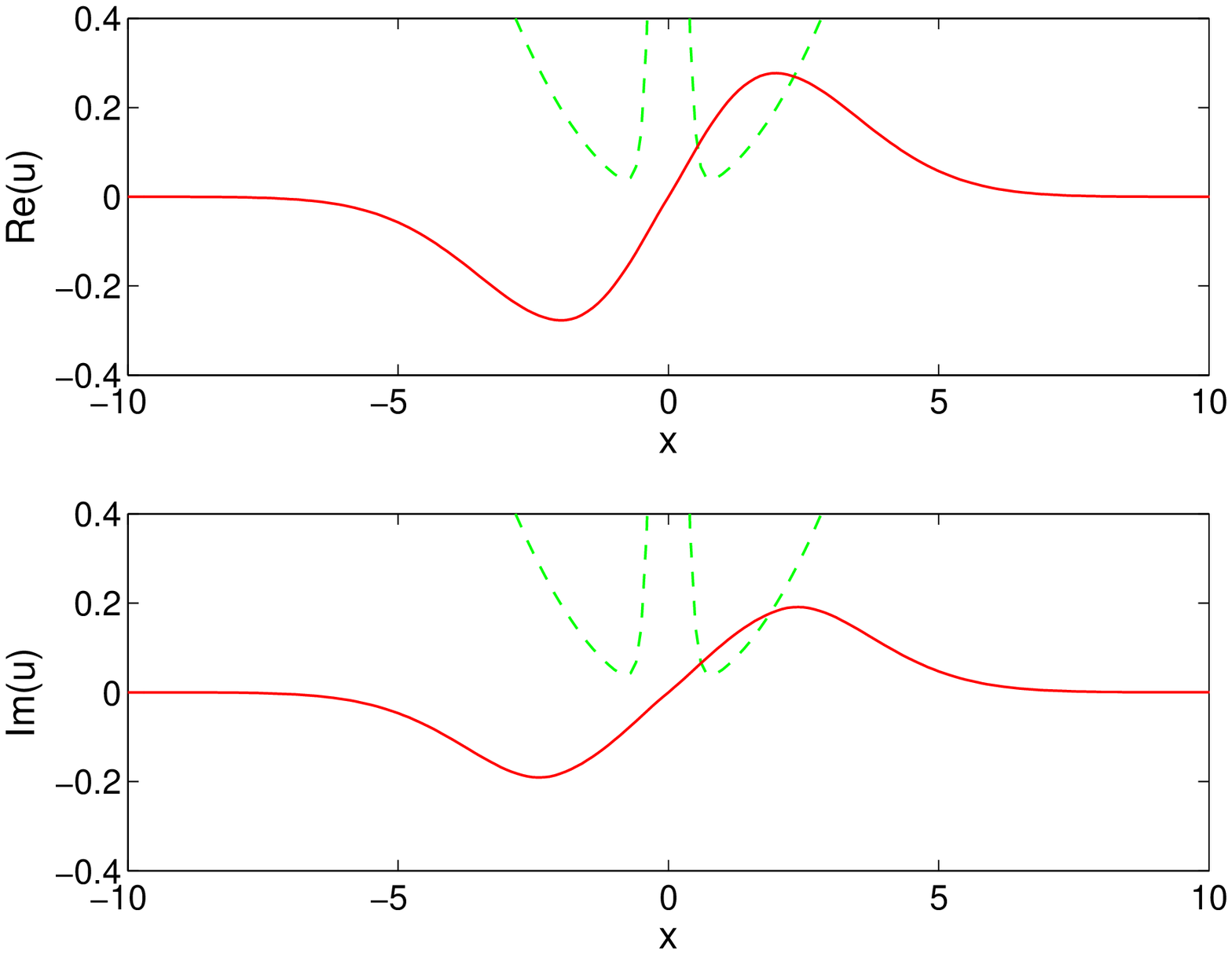} &~~
    \includegraphics[width=8.5cm]{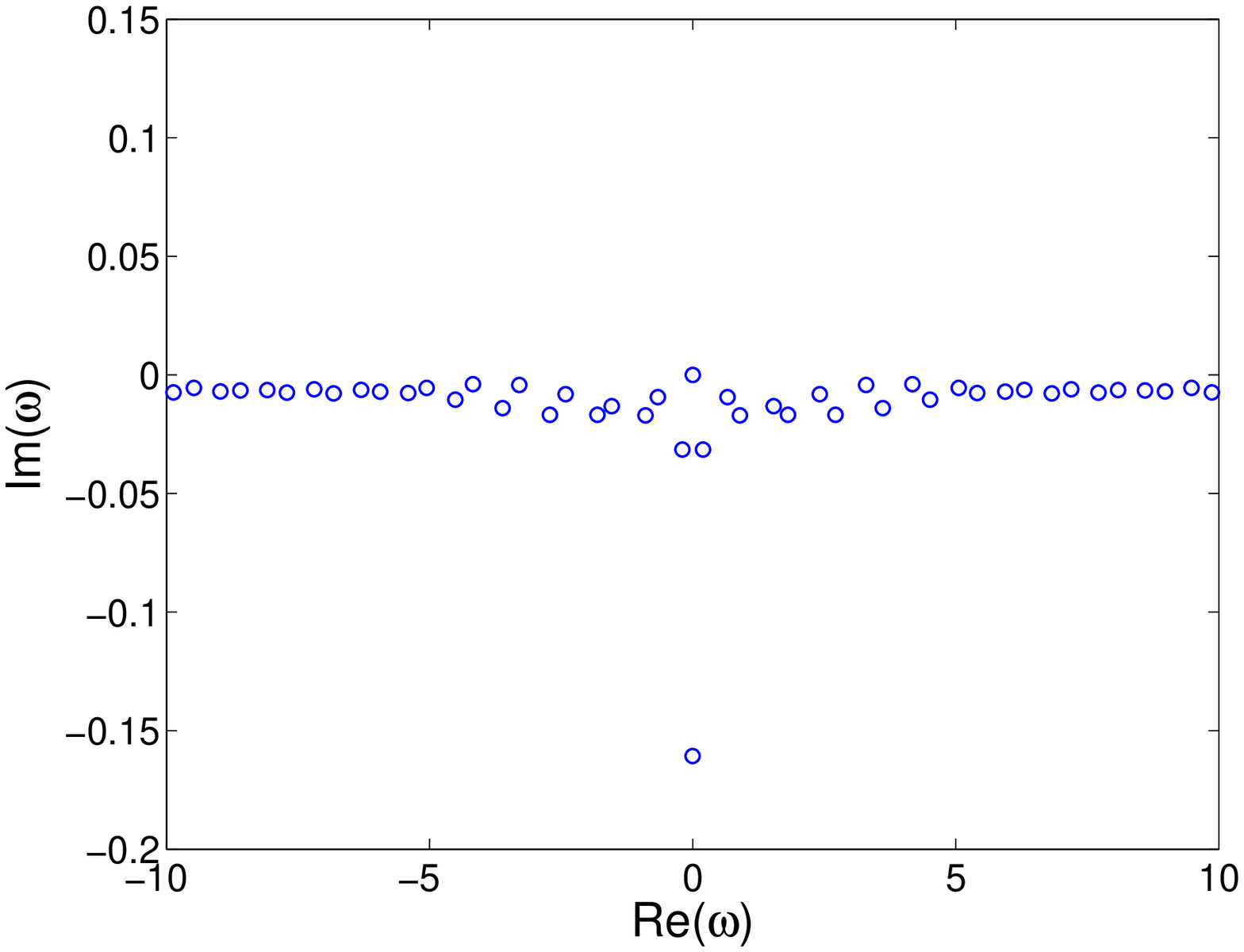} \\
\end{tabular}
\vspace{-0.3cm}
\caption{(Color online) (Left) Real and imaginary part of the wavefunction profile
for a symmetric (top) and anti-symmetric (bottom) solution. (Right) Their corresponding stability eigenvalues. In all cases $\sigma=1$ and 
the nonlinearity is attractive ($s=-1$).}
\label{fig:profatt1}
\end{center}
\end{figure}

\begin{figure}[htpb]
\begin{center}
\begin{tabular}{cc}
    \includegraphics[width=8.2cm]{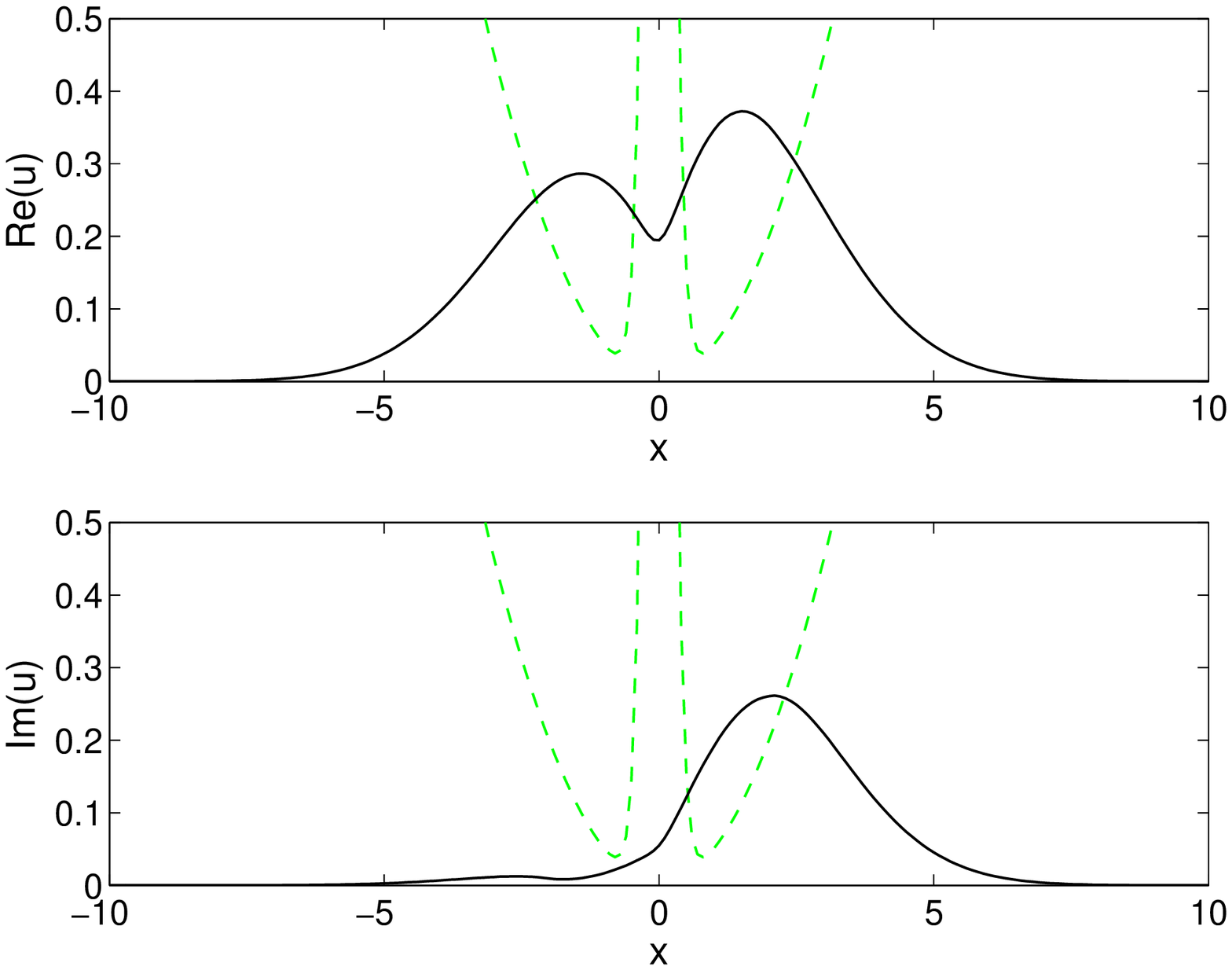} &~~
    \includegraphics[width=8.5cm]{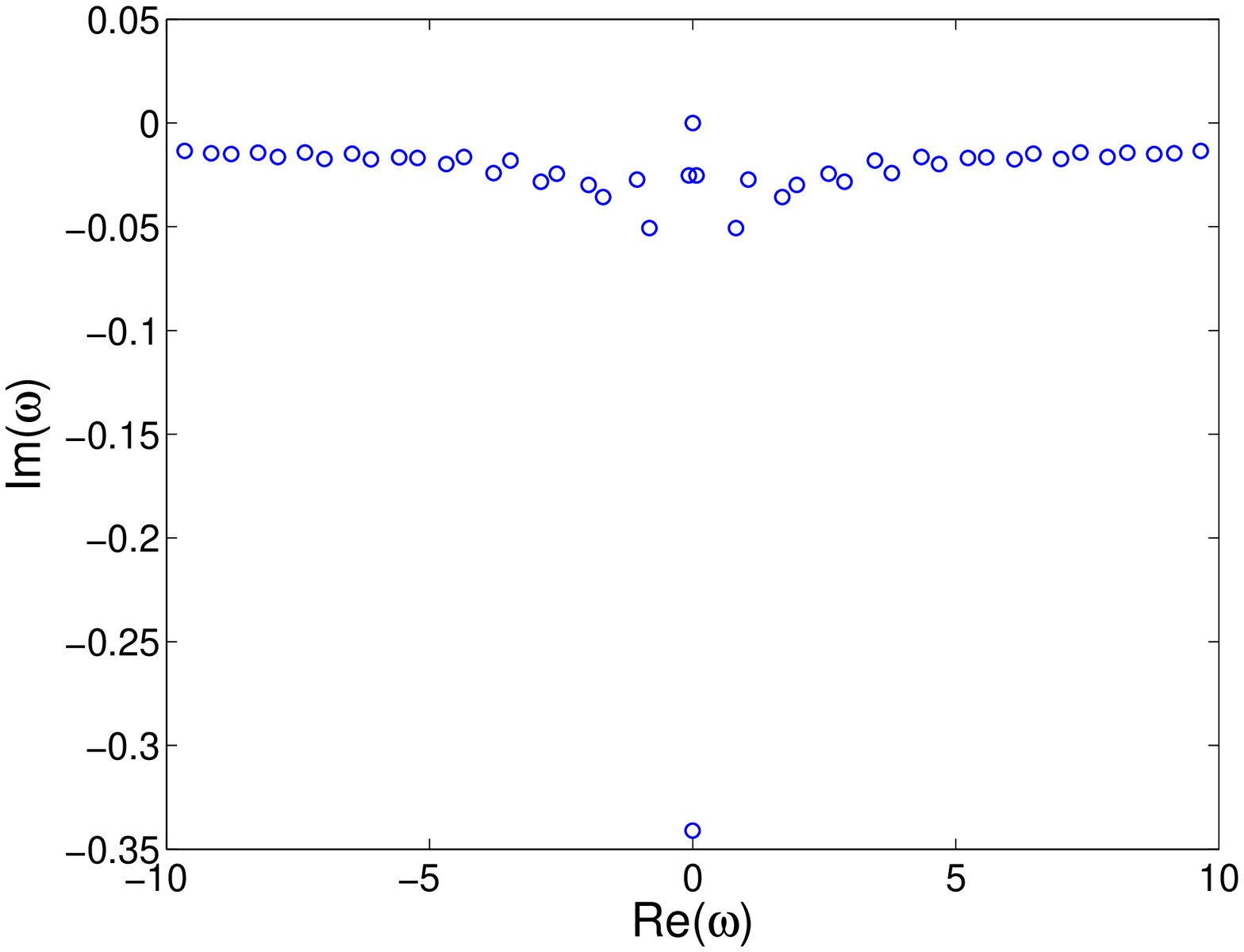} \\
    \includegraphics[width=8.2cm]{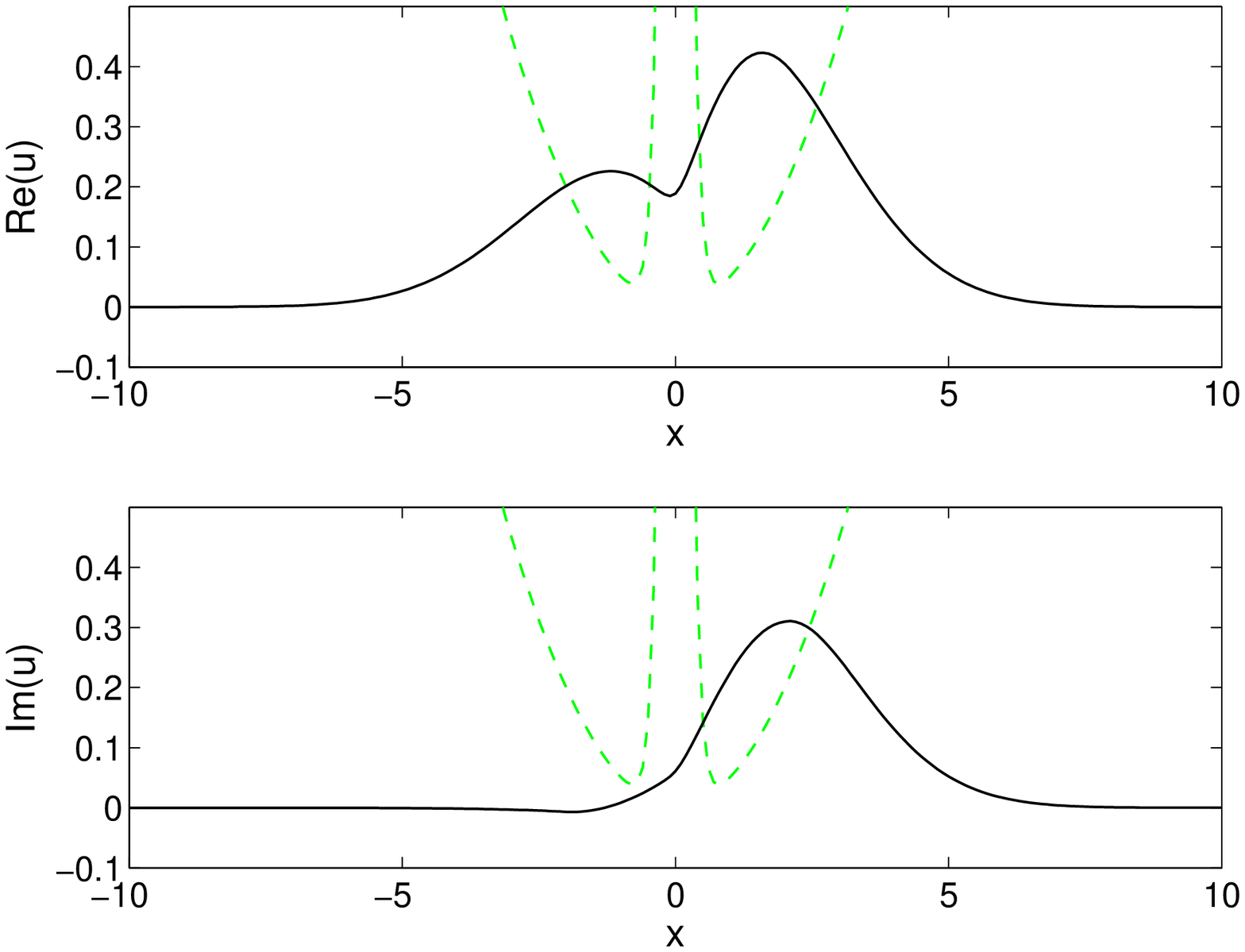} &~~
    \includegraphics[width=8.5cm]{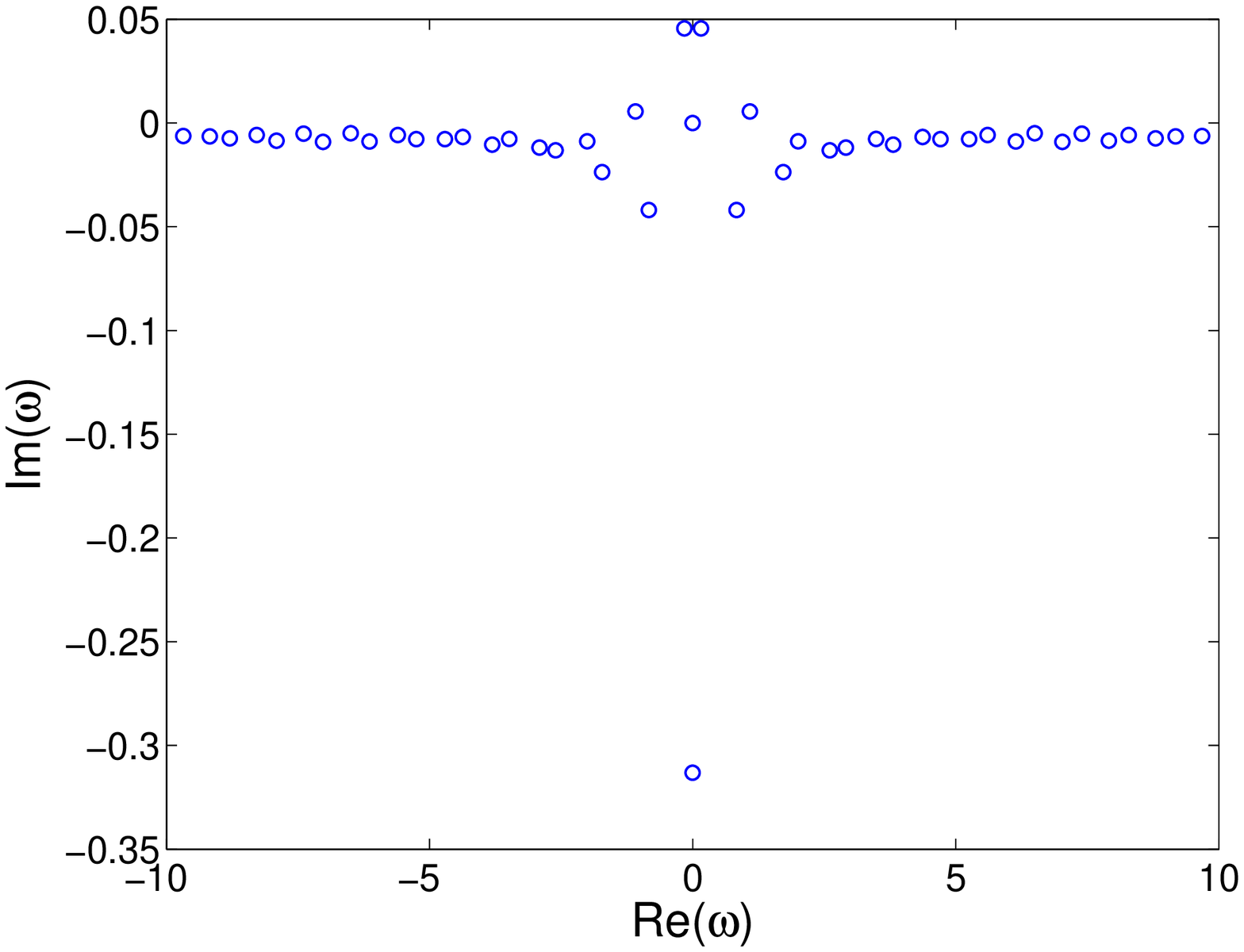} \\
\end{tabular}
\vspace{-0.3cm}
\caption{(Color online) (Left) Real and imaginary part of the wavefunction profile for an asymmetric solution with $\sigma=1$ (top) and $\sigma=0.7$ (bottom). (Right) Their corresponding stability eigenvalues. Notice the Hopf bifurcations
and the associated oscillatory instabilities through two complex pairs which
have occurred in the latter case. Here, the nonlinearity is attractive ($s=-1$).}
\label{fig:profatt2}
\end{center}
\end{figure}

\begin{figure}[htpb]
\begin{center}
\begin{tabular}{cc}
    \includegraphics[width=8.5cm]{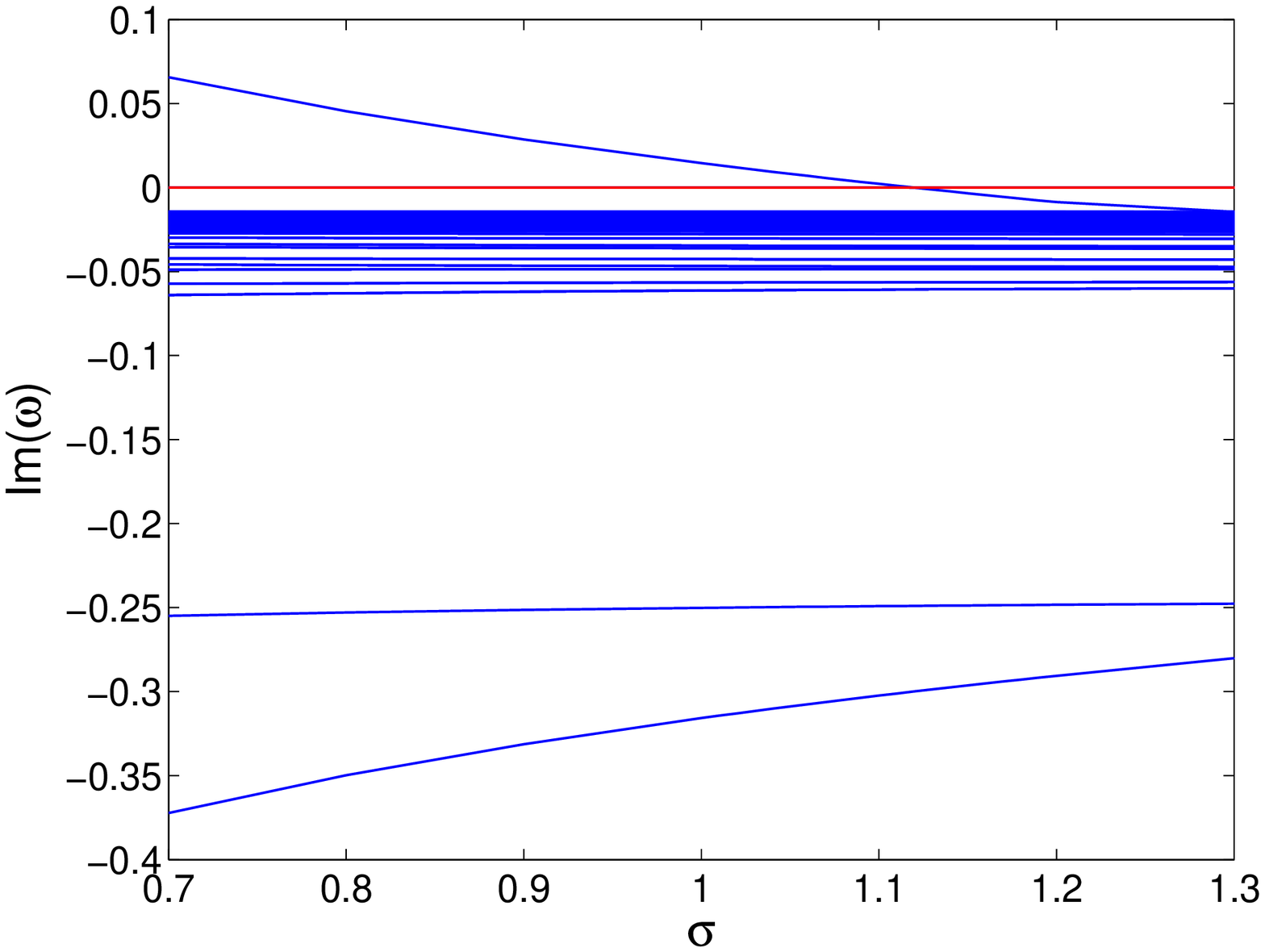} &
    \includegraphics[width=8.3cm]{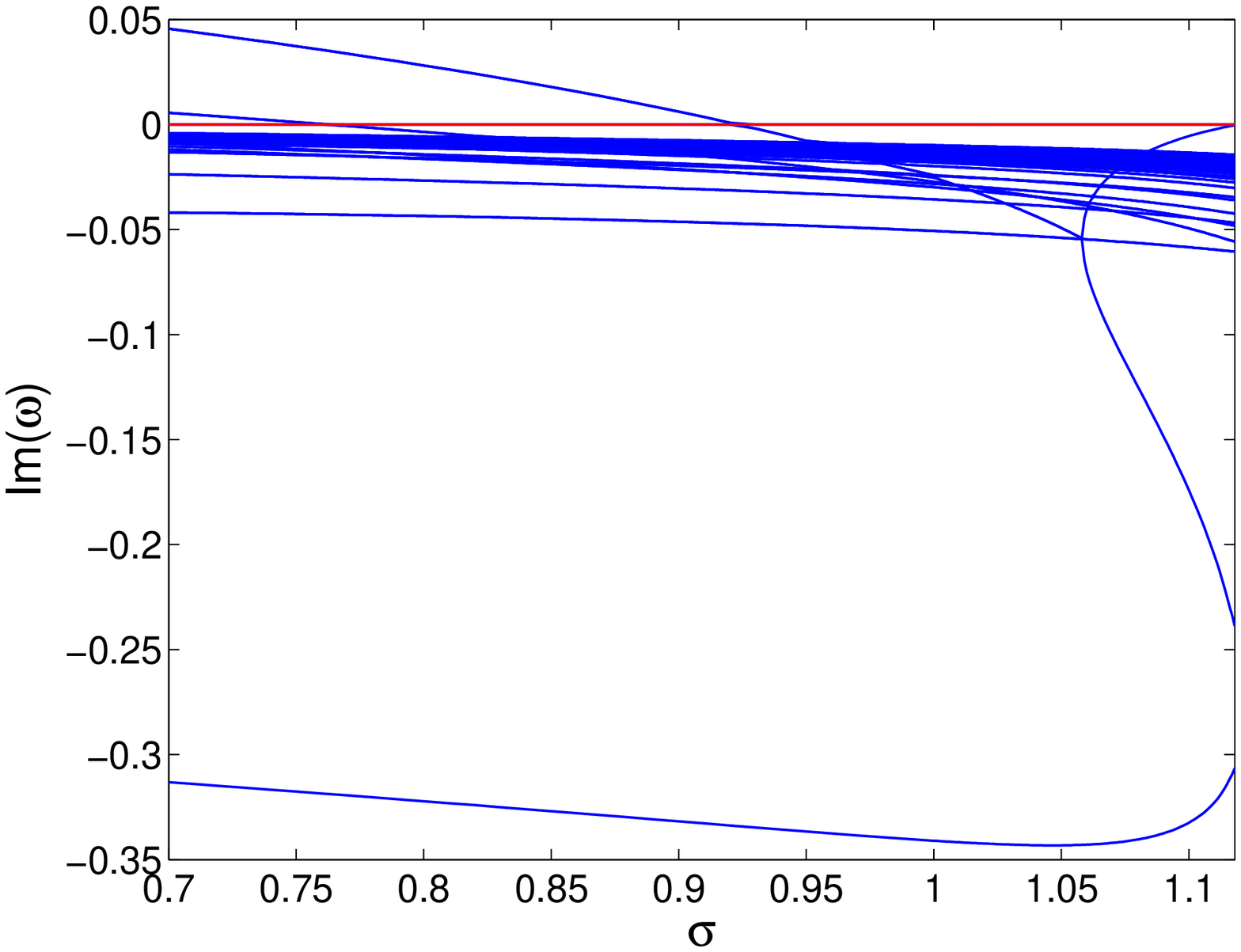} \\
\end{tabular}
\vspace{-0.3cm}
\caption{(Color online) Dependence of the imaginary part of the stability eigenvalues with respect to $\sigma$ for symmetric (left) and asymmetric solutions (right). Here, again, the nonlinearity is attractive ($s=-1$).}
\label{fig:stabatt}
\end{center}
\end{figure}

\begin{figure}[htpb]
\begin{center}
\begin{tabular}{cc}
    \includegraphics[width=8.25cm,height=5.6cm]{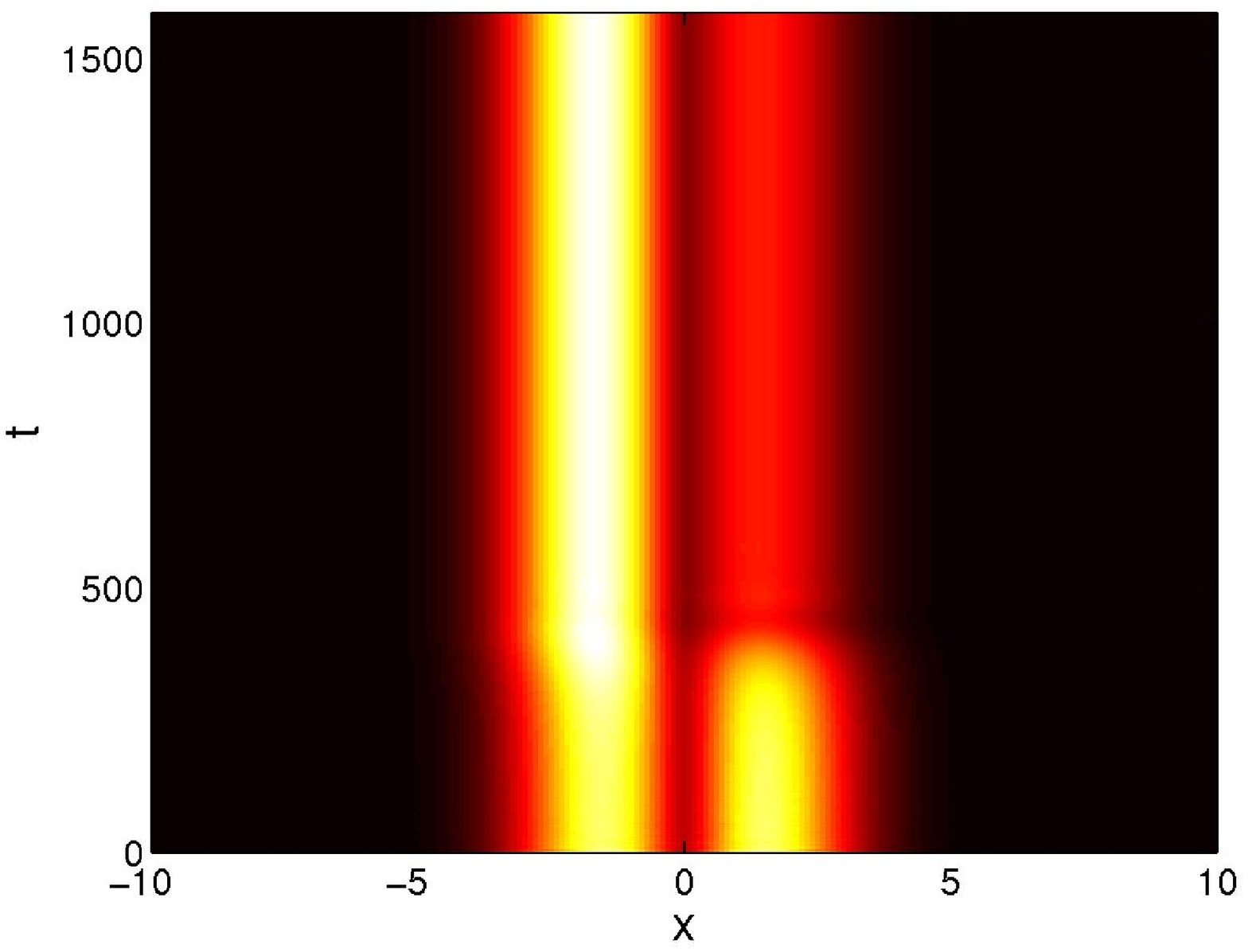} &
    \includegraphics[width=8.25cm,height=5.6cm]{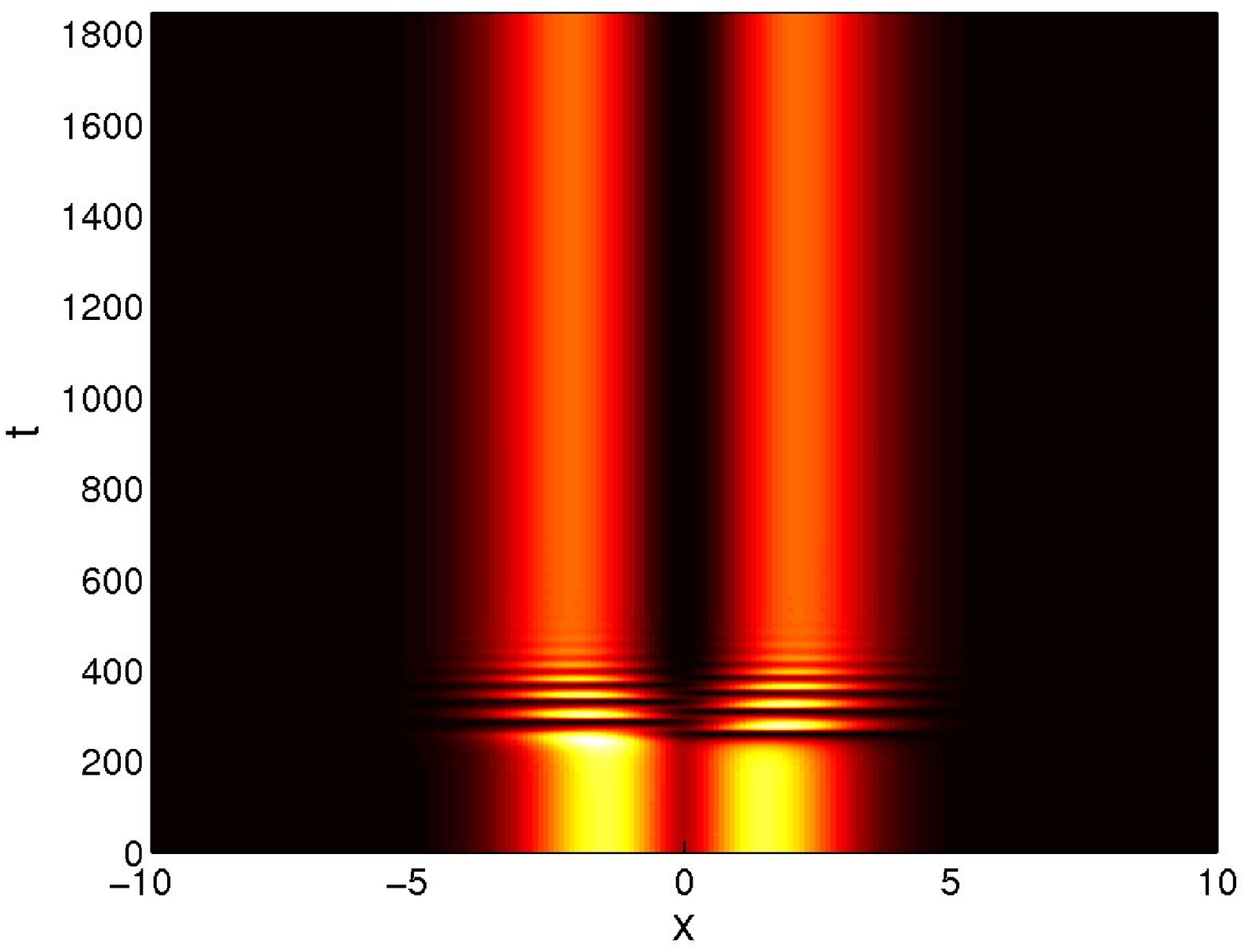} \\
\end{tabular}
    \includegraphics[width=17.2cm]{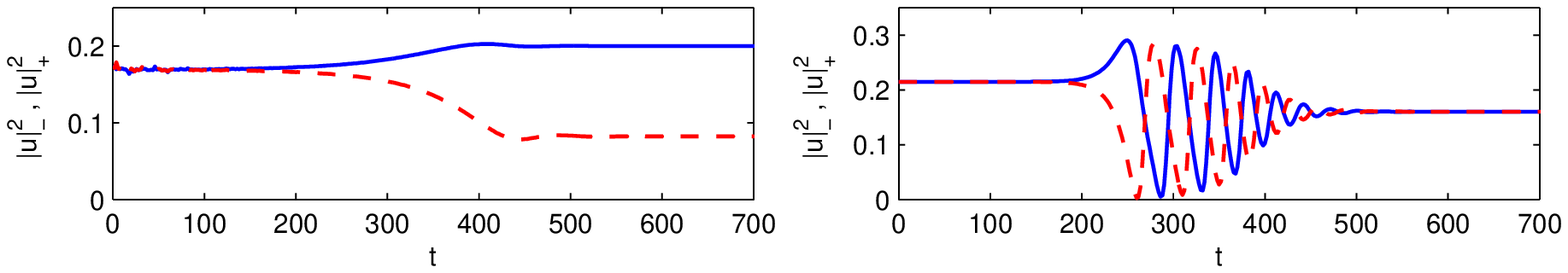}~
\vspace{-0.3cm}
\caption{(Color online) Top: Evolution of perturbed symmetric solitons with
$\sigma=1$ (left) and $\sigma=0.82$ (right) for
attractive nonlinearity ($s=-1$).
Bottom: Respective time
series for the density at the bottom of the left (solid blue line) and right
(dashed red line) wells.
}
\label{fig:dynatt}
\end{center}
\end{figure}

\begin{figure}[htpb]
\begin{center}
\begin{tabular}{cc}
    \includegraphics[width=8.25cm,height=5.6cm]{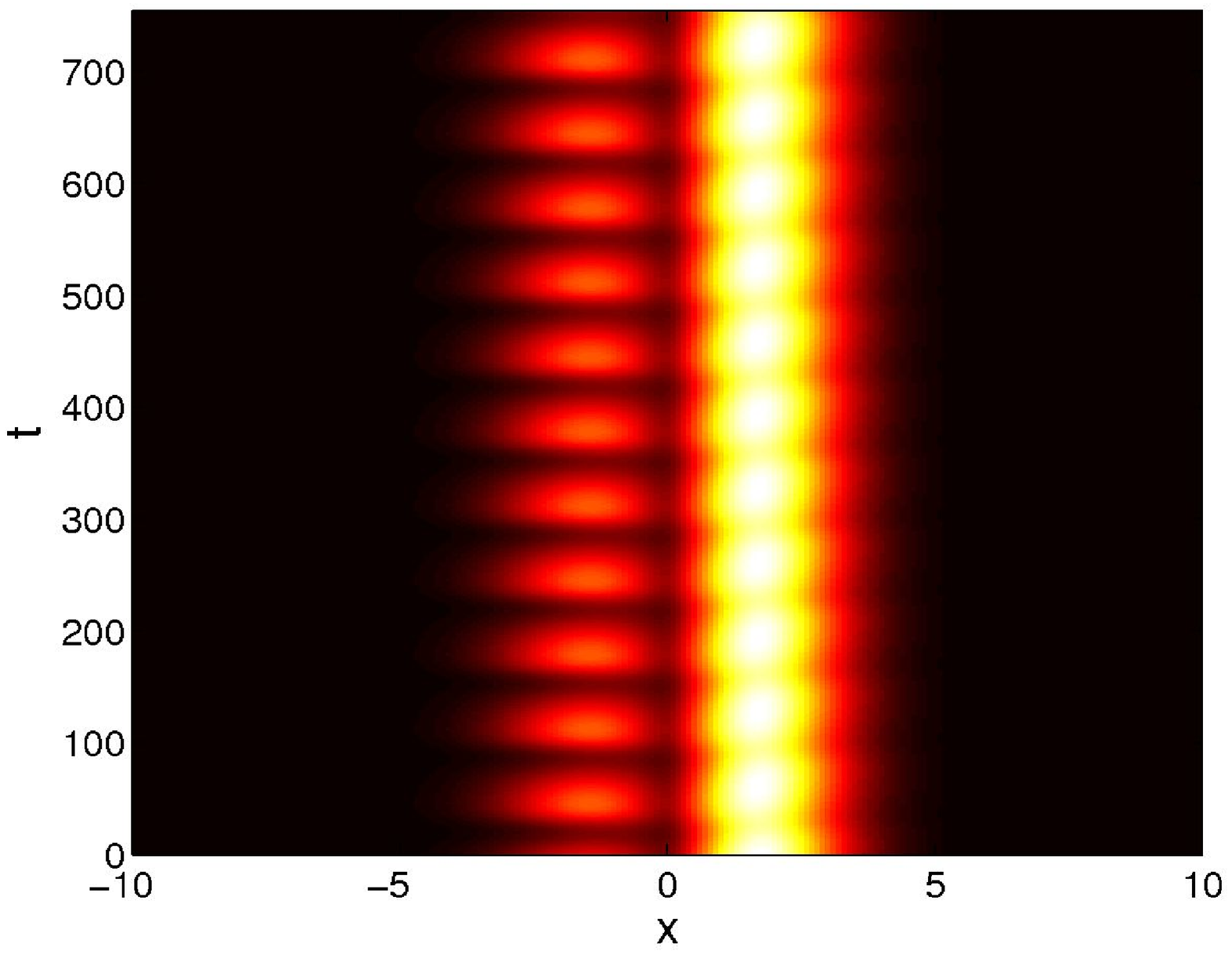} &
    \includegraphics[width=8.25cm,height=5.6cm]{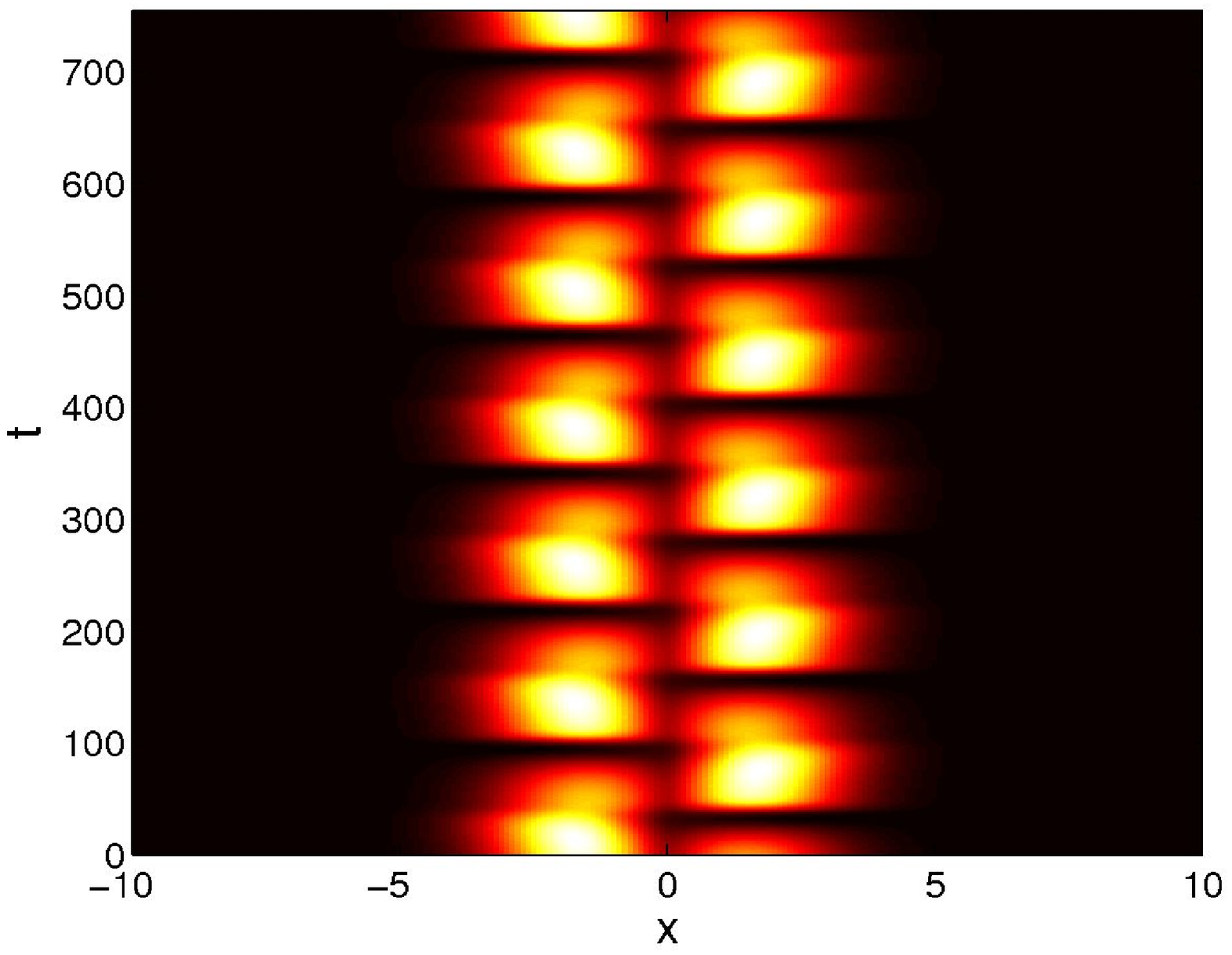} \\
\end{tabular}
    \includegraphics[width=17.2cm]{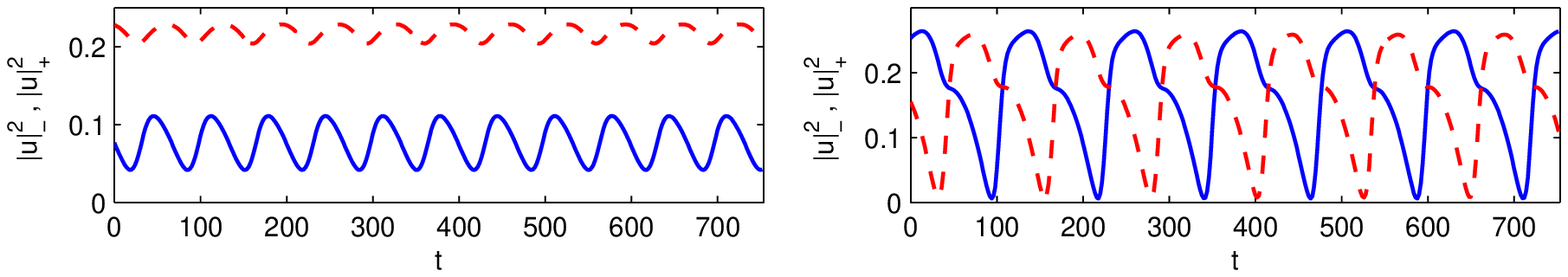}~~~
\vspace{-0.3cm}
\caption{(Color online) Top: Dynamical evolution of the density of the non-stationary asymmetric solution branch found for attractive
nonlinearity ($s=-1$) in two cases: 
$\sigma=0.9$ (left) and $\sigma=0.8$ (right).
Bottom: Respective time
series for the density at the bottom of the left (solid blue line) and right
(dashed red line) wells.
}
\label{fig:dynattnew}
\end{center}
\end{figure}

We have performed a continuation of symmetric, anti-symmetric and asymmetric
states in both cases of repulsive and attractive interactions.
The continuations have been performed by increasing the damping
parameter $\sigma$, which is tantamount to decreasing the
norm or chemical potential. It is important to note that the
chemical potential is no longer a free
parameter in the present setting in
sharp contrast to what is the
case in the Hamiltonian regime of atomic BECs (see also the discussion
of Refs.~\cite{berloff1,augusto}). Similar results can be obtained by decreasing
the pumping parameter $\alpha$. However,
a crucial realization that emerges from considering variations of the
different parameters is that
the spot size $x_m$ must be chosen in a very limited range in order for
the three above mentioned nonlinear modes to co-exist and be potentially
stable; outside this range, instabilities lead to breathing multi-bump
coherent structures. In what follows, the values of
 $x_m=2$ and $\alpha=0.2$ have been used unless explicitly indicated
otherwise.

\subsection{Repulsive case}

We start by considering the case of the
repulsive interaction with $s=+1$ (and vary $\sigma$ as mentioned above).
The family of symmetric solutions is found to be always stable.
As expected, on the other hand, and in agreement to our expectation
from the realm of atomic BECs, the anti-symmetric solutions are
exponentially unstable for small $\sigma$, which is tantamount
to large polariton population numbers $N$. They become stable after the
symmetry-breaking pitchfork bifurcation occurring
at $\sigma=1.045$ (i.e., for $\mu<\mu_{cr}=0.7574$ and for $N<N_{cr}=0.5333$).
The asymmetric branch that emerges through this bifurcation
is stable for $\mu<0.7603$ and $N<0.5509$, i.e., for a narrow
parametric interval past the bifurcation critical point.
However, past this secondary critical point, the asymmetric solutions
are prone towards an oscillatory instability emerging through
a Hopf bifurcation (the critical loss strength in this case is
$\sigma=0.989$). The relevant bifurcation diagrams are presented in
Fig.~\ref{fig:chemrep}, which shows the dependence of $\mu$ on
$\sigma$, as well as the dependence
of $N$ on $\mu$ (note that the latter form of the bifurcation diagram is more
commonly used in relevant studies).
%
The latter graph also contains the results of the theoretical analysis
for the symmetric branch of Eq.~(\ref{eq:ss_sol1}) and for the anti-symmetric
one of Eq.~(\ref{eq:ss_sol0}), as well as for the asymmetric branch
which is theoretically predicted for the parameters of our double-well potential
to bifurcate from the anti-symmetric solution for $\mu>0.7722$ and $N>0.6661$. As can
be seen (also from Fig.~\ref{fig:chemrep}), there is good agreement
between theoretical predictions and numerical findings.

Some case examples of solution profiles for the different branches,
together with the results of their corresponding linear stability
analysis as performed by means of the Bogolyubov-de Gennes (BdG)
ansatz~\cite{emergent} are shown in Figs.~\ref{fig:profrep1} and
\ref{fig:profrep2}, The BdG analysis is represented by the spectral
plane of the linearization eigenfrequencies $\omega={\rm Re}(\omega) + i
{\rm Im}(\omega)$. Contrary to what is the case in the Hamiltonian setting
of Ref.~\cite{TKF06} (where the spectrum is chiefly on the imaginary
axis), here the spectrum contains predominantly decaying modes
with ${\rm Im}(\omega)<0$. For the stable symmetric ground state in
Fig.~\ref{fig:profrep1}, all modes are decaying except for
the symmetry mode associated with $\omega=0$, while for the
unstable anti-symmetric mode of the bottom panel the eigenfrequency
associated with the growth is purely imaginary with ${\rm Im}(\omega)>0$.
On the other hand, for the asymmetric modes of Fig.~\ref{fig:profrep2},
it is evident that shortly past the critical point for their emergence,
a genuine (now that the system is dissipative, in nature) Hopf bifurcation
arises through the crossing of a complex conjugate pair through
the axis of ${\rm Im}(\omega)=0$. Additional Hopf bifurcations happen for
smaller values of $\sigma$ (larger values of $N$), a case example of
which is evident in the bottom panel of Fig.~\ref{fig:profrep2}.
The dependence of the imaginary part of the relevant eigenvalues for the
anti-symmetric and asymmetric solutions with respect to $\sigma$ is
shown in Fig.~\ref{fig:stabrep}, illustrating, respectively, the
relevant pitchfork (left panel) and multiple Hopf bifurcations (right
panel). Naturally, the Hopf bifurcation of the asymmetric branch is anticipated
to give rise to a limit cycle attractor within the dynamics [the relevant
solution is expected to be periodic in the squared modulus of the wavefunction,
hence quasi-periodic in the original field $u(x,t)$].

Two examples of the dynamics of unstable anti-symmetric solutions are illustrated in Fig.~\ref{fig:dynrepanti}.
It is observed that the unstable solutions generically tend to the stable attractors. However, interestingly,
in the  $\sigma=1$ case, the attractor of relevance consists of an
asymmetric steady state, while in the $\sigma=0.8$ case it consists of a symmetric one (the ground state of the system).
%
The symmetry and asymmetry of the configurations can be easily seen
from the time series of the densities $|u|_-^2$ and $|u|_+^2$ measured,
respectively, at the bottom of the left and right wells. These time series
are depicted in the lower panels of the figure.
%
The relevance of the asymmetric attractor, especially for larger values
of $N$ (smaller values of $\sigma$, where the only stable steady
state is the symmetric one) is
confirmed by the simulation shown in the left panel of Fig.~\ref{fig:dynrepasym},
where the dynamics of an unstable asymmetric solution is traced,
leading indeed to the same attractor. The right panel of
Fig.~\ref{fig:dynrepasym} shows the evolution
of a perturbed asymmetric state close to the Hopf bifurcation; in that case, it is observed
that the soliton relaxes to a quasi-periodic asymmetric solution.
[Recall that these solutions have a quasi-periodic evolution for the wavefunction (due to
the periodic evolution of the phase through $e^{-i\mu t}$) but
the evolution in density is {\em periodic} as the panels show]. That quasi-periodic branch
is observed to exist in the range $\sigma\in[0.970,0.990]$.

\subsection{Attractive case}

In the case of attractive interactions ($s=-1$), the scenario is similar
in nature, except for the origin of the symmetry breaking bifurcation.
More specifically, now, the asymmetric solutions, which stabilizes at
$\sigma=0.923$ ($\mu=0.4182$ and $N=0.7199$), bifurcates from
the symmetric solutions branch at $\sigma=1.118$ ($\mu=0.4101$ and $N=0.7741$).
Figures~\ref{fig:chematt} and \ref{fig:stabatt} are the equivalent to
Figs.~\ref{fig:chemrep} and \ref{fig:stabrep}, respectively, but for $s=-1$. Nevertheless, we observe
that both the dependence of the chemical potential $\mu$ on the
nonlinear saturation parameter $\sigma$ and that of $N$ on $\mu$
is, in fact, non-monotonic for this example in the case
of the bifurcating asymmetric branch. This clearly indicates
(see the right panel of Fig.~\ref{fig:chematt}) that the relevant
bifurcation is {\it subcritical} (as the chemical potential
$\mu$ is decreased, which is the natural direction of
variation off of the linear limit). This is contrary to the corresponding
supercritical expectation of its Hamiltonian analog~\cite{TKF06,kirr}.
It should be noticed, however, that other examples where such subcritical
bifurcations have been previously reported
in Refs.~\cite{sacch,kirr2} although in neither case was the nonlinearity
purely cubic as was the case here (and they did not contain driving/damping
effects). Importantly, it should also be pointed out that the
analytical prediction of the Galerkin approach suggests a supercritical
scenario for $\mu<0.4247$ and $N>0.6590$. Despite the inability of the
approximation to capture the short subcritical segment of the bifurcating
branch, we nevertheless see that the Galerkin method is a useful tool
for obtaining an estimate of the relevant critical point.

An additional feature worth pointing out concerns the nature of the
instabilities of the different branches as detailed in
Figs.~\ref{fig:profatt1} and \ref{fig:profatt2}. While the symmetric
branch becomes unstable at the relevant critical point by developing
an imaginary eigenfrequency with ${\rm Im}(\omega)>0$
(the rest of the spectrum has ${\rm Im}(\omega)<0$),
the anti-symmetric state remains dynamically robust. On the other hand,
the asymmetric branch emerges as stable at the critical point of the
symmetry breaking but shortly thereafter (for $\sigma<0.923$), it becomes
subject to a Hopf bifurcation through the crossing of the axis with
${\rm Im}(\omega)=0$ of a complex eigenvalue pair.
In fact, for $\sigma< 0.74$, a secondary Hopf bifurcation has
occurred and is mirrored in the two complex pairs with ${\rm Im}(\omega)>0$
shown in Fig.~\ref{fig:profatt2}. This phenomenology is enforced
by Fig.~\ref{fig:stabatt} which illustrates the dependence of the relevant
stability eigenvalues on the nonlinear loss parameter $\sigma$ (see the
right panel for the sequence of Hopf bifurcations, while the left panel
highlights the symmetry-breaking induced
crossing of a single eigenfrequency pair for the symmetric branch).
As in the repulsive case, the Hopf bifurcation of the asymmetric branch is anticipated
to give rise to a limit cycle attractor within the dynamics.

The dynamics of Figs.~\ref{fig:dynatt} and \ref{fig:dynattnew} naturally
reflects the above conclusions. In particular, the evolution of the symmetric
state in the double-well potential of the left panel of Fig.~\ref{fig:dynatt} gives
rise to the asymmetric state as the latter is stable and indeed an
attractor
for the value of $\sigma=1$. The right panel of the figure displays the evolution
of a perturbed symmetric solution tending to an anti-symmetric one; in that case,
the asymmetric solution is unstable and no longer a dynamical attractor.

On the other hand, Fig.~\ref{fig:dynattnew} shows different case examples
of the (unstable via the Hopf) asymmetric branch for different values of
$\sigma$. In those cases, the asymmetric branch is no longer
a stable stationary state and as a result the dynamics becomes
periodic in the modulus (quasi-periodic in the original field)
for $\sigma\in[0.74,0.92]$.
It is interesting to follow the changes in the dynamics for these periodic
states as $\sigma$ is decreased below the bifurcating point from the
asymmetric branch. In particular, close to bifurcation point, the periodic
evolution remains proximal to the state from which it emanates, namely the
asymmetric state as it can be seen in the left panels of Fig.~13. However,
as $\sigma$ is decreased further from the bifurcation point, the instability
of the asymmetric state is stronger and the departure from the asymmetric
solution is more significant. In particular, it is interesting to notice
that for smaller values of $\sigma$, the solution tends to display strong
oscillations of the densities resembling the {\em symmetric} tunneling of
matter from one well to the other. An example of this evolution for
$\sigma=0.8$ is depicted in the right panels of Fig.~13 where it is
evident that the oscillations in the two wells become similar to each
other but with a phase shift between them, leading to an effective
re-symmetrization of the dynamics.

It is also interesting to highlight here the difference between the
repulsive case of Figs.~\ref{fig:dynrepanti} and \ref{fig:dynrepasym}
and the attractive case of Figs.~\ref{fig:dynatt} and \ref{fig:dynattnew}.
In the former case, when the emerging asymmetric branch is unstable
the dynamics typically is found to lead to the stable ground state of
the system (the symmetric one). On the other hand, for the attractive case,
when both the symmetric and the asymmetric branch are destabilized, the
dynamics does not resort to the excited (yet stable) anti-symmetric state.
Instead, it leads to periodic oscillations in the density between the
two wells.


Finally, we have considered the effect of varying the spot size fixing $\sigma=1$. In the repulsive case, the symmetric branch is stable for $x_m\in[0.9,5.7]$; out of this range, the instabilities are caused by a Hopf bifurcation cascade and develop into non-stationary multi-dark soliton waveforms, similar to the states that
were previously reported in Ref.~\cite{augusto} (but for a purely parabolic trap).
The anti-symmetric branch, which is unstable for every $x_m$ (for this value
of $\sigma$), experiences a bifurcation cascade for $x_m\leq2.0$ and $x_m\geq5.3$. The instabilities at $x_m\in(2.0,5.3)$ are the exponential ones
previously explored. However, considering higher values of $\sigma$,
 a stability range appears which is enlarged for growing $\sigma$. A
similar effect is observed for the asymmetric branch, i.e., there is a small
stability interval $x_m\in[1.9,2.0]$ that is enlarged
when $\sigma$ is decreased. Outside this range, the branch experiences
Hopf bifurcation cascades.

The above mentioned scenario is almost equivalent for the attractive case,
except that the symmetric and anti-symmetric branches are interchanged. In
that case, the anti-symmetric branch is stable for $x_m\in[2.0,4.8]$; the
symmetric branch is now stable for $x_m\in[1.0,1.9]$, starting the Hopf
cascade at $x_m=4.5$. The asymmetric branch is stable for $x_m\in[1.0,2.0]$,
while being oscillatorily unstable for other values of $x_m$.

\section{Conclusions and Future Challenges}

In the present work, we
studied the existence of solutions,
their spectral stability and nonlinear dynamics for the case of
a polariton
condensate confined in a quasi-1D double well potential.
Motivated by recent developments for the study of polaritons
in such settings
\cite{1d_polaritons1,1d_polaritons2,1d_polaritons3,1d_polaritons4,1d_polaritons5,1d_polaritons6,Amo:10,polardw},
and by the work of Ref.~\cite{polar1} which proposed a two-well model,
we presented a systematic
Galerkin analysis for the model with the gain over a localized
spot and nonlinear saturation loss formulated
in Refs.~\cite{berloff1,kbb,review_nb}.
It was theoretically predicted that nonlinear states emanate from the
corresponding linear ones of the potential and that bifurcations
are expected to arise, similarly to the Hamiltonian analog of this
setting studied earlier in the context of atomic BECs. Such symmetry breaking
pitchfork events emerge from the anti-symmetric, first excited
state in the case of the repulsive interactions, while they
arise from the symmetric ground state branch in the case of
attractive ones. Despite the similarities with the atomic BEC
case, nontrivial differences exist as well. One of them concerns
the nature of the bifurcation, which in the attractive case was
found to be weakly subcritical (instead of supercritical) upon decrease
of the chemical potential. Importantly also, the resulting asymmetric
branches aside from narrow intervals of stability are generically
found to be unstable due to genuine Hopf bifurcations, which, in turn,
give rise to periodic orbits (in the density). While in the repulsive
case, the dynamics of anti-symmetric and asymmetric branches is found
to be attracted to the ground state when both of them are unstable,
the periodic orbits are essential to the evolution in the case of
attractive interactions as they seem to constitute the robust
dynamical attractor.

This is merely the first step in the examination of the
similarities (but also the differences) of the polariton BECs
and their atomic counterparts within a setting that contains the
interplay of a double-well potential and nonlinear interactions.
Yet, our study paves the way for a number of potential future avenues.
On the one hand, one can consider the more detailed model of
Refs.~\cite{polar1,polar2,cc05} and examine whether the inclusion of
the diffusive dynamics of the exciton population induces any
qualitative differences in the features reported herein.
On the other hand, and bearing in mind the predominantly two-dimensional
nature of the polariton dynamics, one can envision generalizations of
the potential considered herein in a 2D realm. Relevant possibilities
may include not only the straightforward generalization of a double
well encompassing two quasi-one-dimensional tracks, but also that of
a genuinely two-dimensional four well potential that has recently
been examined in detail in atomic BECs~\cite{chenyu_2d}.
Even in the context of the present model, there are further possibilities
to explore, including the systematic investigation of the emergent
periodic orbits and their Floquet spectral stability analysis.
Such studies
are currently in progress and will be reported in future publications.

\section*{Acknowledgments}
%
%
J.C. acknowledges financial support from the MICINN project FIS2008-04848.
P.G.K.~and R.C.G.~gratefully acknowledge support from the National Science Foundation
under grant DMS-0806762. P.G.K.~also acknowledges support the Alexander
von Humboldt and Binational Science Foundations.
The work of D.J.F.~was partially supported by the
Special Account for Research Grants of the University of Athens.

\end{document}